\documentclass[useAMS,usenatbib]{mn2e}
\usepackage{times}
\usepackage{url}
\usepackage{graphicx}
\usepackage[usenames]{color}
\DeclareGraphicsExtensions{.pdf,.png,.jpg,.mps,.eps,.ps}
\usepackage{amsmath}
\usepackage{natbib}

\usepackage{tikz}

\def\unit #1{\,{\rm #1}}

\newcommand\cmsqi{\rm \,\unit{cm^{-2}}}

\newcommand\kev{\rm \,\unit{keV}}

\newcommand\funit{\rm \,erg\,cm^{-2}\,s^{-1}}

\newcommand\lunit{\rm \,erg \,s^{-1}}

\newcommand\xiunit{\rm \,erg\,cm\,s^{-1}}

\newcommand\lambdaedd{\lambda_{\rm \, Edd}}

\newcommand\nh{ N_{\rm H}}

\newcommand\pc{\unit{pc}}

\newcommand\kpc{\unit{kpc}}

\newcommand\ev{\unit{\, eV}}

\newcommand\lbol{L_{\rm \, bol}}
\newcommand\ledd{L_{\rm \, Edd}}

\newcommand\lhard{L_{\rm 2-10 \kev}}

\newcommand\lsoft{L_{\rm 0.3-2 \kev}}
\newcommand\luv{L_{\rm 2500 \AA}}

\newcommand\mbh{M_{\rm BH}}
\newcommand\msol{M_{\odot}}

\newcommand\chandra{{\it Chandra}}

\newcommand\xmm{{\it XMM-Newton}}

\newcommand\suzaku{{\it Suzaku}}
\newcommand\nustar{{\it Nustar}}

\usepackage{graphicx}

\title{An X-ray view of central engines of low luminosity quasars (LLQSO) in the local Universe.}
\author[Laha et al.]{Sibasish Laha$^{1}$\thanks{sib.laha@gmail.com, slaha@uscd.edu}, Ritesh Ghosh$^2$, Matteo Guainazzi$^{3}$ \thanks{Matteo.Guainazzi@sciops.esa.int} and Alex G. Markowitz $^{4,1}$. \\
$^1$ University of California, San Diego, Center for Astrophysics and Space Sciences, 9500 Gilman Dr, La Jolla, CA 92093-0424, USA.\\
$^2$Visva-Bharati University, Santiniketan, Bolpur 731235, West Bengal, India. \\
$^{3}$European Space Research and Technology Centre, Keplerlaan 1, 2201 AZ Noordwijk, Netherlands.\\
$^{4}$Nicolaus Copernicus Astronomical Center, Polish Academy of Sciences, Bartycka 18, PL-00-716 Warszawa, Poland.\\
}
\date{\today}
\begin{document}
\pagerange{\pageref{firstpage}--\pageref{lastpage}} \pubyear{2018}




\maketitle
\label{firstpage}
\begin{abstract}
	{We have carried out a systematic X-ray spectral analysis of a
        sample of low luminosity quasars (LLQSO) to investigate the
        nature of the central engines of these sources. The optically-selected LLQSO
        sample consists of close, known bright active galactic nuclei
	(AGN) which serves as an important link between the powerful quasars at higher redshift and local Seyfert galaxies. We find that although the bolometric luminosities of the LLQSOs are lower than those of the higher redshift quasars by almost an order of magnitude, their distribution of the Eddington rate $\lambdaedd$ is similar. We detect a strong anti-correlation between $\alpha_{\rm OX}$ and $L_{2500 \rm \AA}$, as has also been detected in several other quasar studies with large sample sizes, indicating that as the UV luminosity of the source increases, the X-ray luminosity decreases. We do not detect any significant neutral obscuration ($\nh\ge10^{22}\cmsqi$) in the X-ray spectra of the LLQSOs, and hence rule out obscuration as a possible cause for their lower luminosity. We conclude that the central engines of the LLQSOs function similarly to those of the higher redshift quasars, and the difference is possibly because of the fact that the LLQSOs have lower black hole masses. We do not find any correlation between the molecular gas in the host galaxies and accretion states of the AGN. This indicates that the presence of molecular gas in the host galaxies of the LLQSOs does not significantly influence the instantaneous accretion rates of their SMBHs.}

\end{abstract}

\begin{keywords}
	galaxies:active, galaxies:evolution, galaxies:kinematics and dynamics, galaxies:quasars:absorption lines, Galaxies:Seyfert.

\end{keywords}

\vspace{0.5cm}


\section{INTRODUCTION}

The co-evolution of super massive black holes (SMBHs) and their host
galaxies is a subject of current active research. Feedback from
Active Galactic Nuclei (AGN) is believed to play a crucial role in
such evolution, and is suspected to be a driver of observed correlations between black hole
mass and the host galaxy bulge velocity dispersion, the $\mbh- \sigma$
relation \citep[][and references
  therein]{2000ApJ...539L..13G,2005SSRv..116..523F,2013ARA&A..51..511K,2016MNRAS.460.3119S,2017MNRAS.466.4029S}.
Observational and theoretical studies \citep[see e.g.,][]{2007MNRAS.382.1415S,2013ApJ...763L..18W}
demonstrate how outflows from AGN may interact with the
host galaxies and may drive away the neutral material responsible for
feeding the black hole. In this scenario, the lack of supply of cold neutral gas in the
vicinity of the SMBH leads to lowered accretion rate and eventually
quenching of AGN activity. It is therefore interesting to study a
sample of sources whose luminosities lie between those of the powerful high
redshift quasars and the low luminosity AGN in the local Universe to
track the evolutionary scenario of quasars and hence the SMBH
host galaxy interaction.

For an SMBH of mass $\sim$ $10^7-10^9\msol$, the primary emission
from the accretion disk peaks in the UV \citep{1973A&A....24..337S}. A
significant fraction of these primary UV photons are upscattered to
X-rays by hot optically-thin gas, popularly known as
the corona \citep[][]{1993ApJ...413..507H,1994ApJ...432L..95H}. The X-ray photons from the corona are reprocessed by circumnuclear
structures to produce several broadband continuum
features such as the soft X-ray excess and the Compton hump, and also discrete
features, namely emission lines in soft X-rays and the narrow and
broad Fe K $\alpha$ emission lines at $\sim 6.4 \kev$. The UV emission in particular contributes to the bulk of the total bolometric luminosity
emitted by the central engine of an AGN. Studying AGNs' UV and X-ray emission
is thus important for diagnosing the activity of the central engine, as
well as for probing the structure of the surrounding reprocessing media.
Additional spectral complexity in the
X-rays arises in the form of narrow absorption features in the soft ($0.3-2\kev$)
and hard X-rays ($6-9\kev$) due to ionised outflows, popularly known as the warm
absorbers and ultra-fast outflows, UFO (See for e.g., \citet{2013MNRAS.430.1102T,2014MNRAS.441.2613L}). These outflows could provide mechanical feedback to
the host galaxy.

The low luminosity QSO (LLQSO) sample
\citep{2007A&A...470..571B,2016A&A...587A.138B} gives us the
opportunity to study the central regions of local quasars that are less luminous than high
redshift quasars (sample selection is described in Section \ref{sec:sample}). \citet{2014A&A...561A.140B} in an extensive near infrared (NIR) study of the LLQSOs found that these sources have lower stellar ($\sim 2\times10^{9}-2\times 10^{11} \msol$) and black hole masses ($\sim 1\times 10^6- 5\times 10^8 \msol$) compared to higher luminosity quasar (QSOs) samples. These sources are also less luminous in terms of both nuclear and host galaxy emission, although \citet{2007A&A...470..571B} have detected large quantities of molecular gas ($\sim 10^9 \msol$) in majority of the LLQSO host galaxies, indicating no dearth of fuel for AGN or star formation activity. The LLQSOs therefore serve as a good sample to study the central AGN properties of lower luminosity local quasars and how they differ from their brighter counterparts at higher redshift, thus shedding light on the evolutionary scenario of AGN. X-rays provide us with direct evidence of nuclear activity, and hence, we study the X-ray spectra of these AGN to answer the questions: 1. How do the X-ray luminosity of the LLQSOs compare with higher redshift quasars; are the X-ray spectra obscured in these sources? 2. Are the LLQSOs inefficiently accreting compared to their higher redshift counterparts even in the presence of large reservoirs of molecular gas or 3. Are these sources scaled down version of brighter quasars with smaller black hole masses, while the AGN physics at their core remains the same?
The LLQSO sources are by selection type 1 AGN which 
gives us a direct view of their central engine. With an unhindered view of the
central engine, the presence of molecular gas in the host galaxy, and
having luminosities lower than the bright quasars, this sample is ideal to test the cause as to why these local
($z<0.06$) quasars are fainter.

In this paper we report the results of a detailed X-ray spectral
analysis of the LLQSOs. The paper is arranged as follows: Section
\ref{sec:sample} describes the LLQSO sample used in this work. Section
\ref{sec:obs} describes the observation and data reduction, Section
\ref{sec:data} describes the steps taken in data analysis. Section
\ref{sec:results} lists the important results of our analysis, while
Section \ref{sec:discussion} discusses the results. This is followed
by conclusions in Section \ref{sec:conclusion}.

\section{Sample description}\label{sec:sample}
\subsection{The LLQSO sample}

The LLQSOs are a subsample of the Hamburg/ESO survey (HES) for bright
quasars \citep{2000A&A...358...77W}. The HES survey consists of 415
bright QSOs and Seyfert 1s which are spectroscopically complete with
respect to flux and redshift limits. The optical magnitudes are in
the range $13\le B_J \le 17.5$, and the redshift range 
is $0 < z < 3.2$. The HES sample has type 1 sources quasars
only. \citet{2014A&A...561A.140B} constructed the LLQSO sample out of
the HES sample, consisting of 99 sources using a redshift cut-off of
$z<0.06$ on the HES sample. This redshift cut-off ensured that the
CO(2-0) band head in the near infra-red (NIR) spectra is detected by the 
IRAM telescope. The LLQSO sample of 99 sources probes the lower
luminosity function tail for the local quasars
\citep[][]{1997A&A...325..502K,2007A&A...470..571B}, and the
redshift-magnitude diagram of the LLQSOs \citep{2014A&A...561A.140B}
shows that the sources lie below the commonly used division line of
$M_{\rm B} =-21.5+ 5 \log h_0$ between QSOs and Seyferts. These
  sources are therefore ideal for studying the evolution between
  bright QSOs and lower luminosity Seyfert galaxies. Out of these 99
sources, only 16 sources have broadband X-ray observations by
\chandra{}, \xmm{} or \suzaku{} and have publicly available
data. This sample of 16 sources is used in this work and will be
referred to as LLQSOs. Table \ref{Table:sources} lists the basic
properties of the sources.

\subsection{The comparison samples}\label{subsec:controlsamples}

 In this work we compare the UV/X-ray spectral properties of the LLQSOs with AGN at various redshift ranges to derive clues on the evolution of the nature of the central engines from high redshift quasars to the local Seyfert galaxies. The most important physical parameters in our study are the bolometric luminosity $\lbol$, the black hole mass $\mbh$, the Eddington rate ($\lambda_{\rm Edd}=\lbol/\ledd$), the $2500\rm \AA$ monochromatic luminosity $L_{\rm 2500 \AA}$, the X-ray luminosity $\lhard$, and the UV to X-ray spectral slope $\alpha_{\rm OX}$. We 
have carried out this comparison by selecting four well studied samples (with publicly available information) spanning different 
redshift ranges: 1) Local Seyfert galaxies, `The Warm Absorber in X-rays, WAX sample',
\citet{2014MNRAS.441.2613L}. This sample consists of 26
 Seyfert 1 galaxies in the local Universe ($z<0.06$);  2) `Palomar Green, PG
quasars', \citet{1994ApJ...435..611L}, consists of optically selected sample of 23 bright quasars
in the near redshift range $z=0.06-1.72$, and 3) the XMM-COSMOS
sample \citep{2010A&A...512A..34L,2014ApJ...781..105L,2016A&A...590A..80R} in the intermediate to far redshift range $z=0.1-3.5$; and 4) the WISSH quasar sample \citep{2017A&A...608A..51M} at far redshift $z=2-4$. The WISSH quasars represent an extreme class of AGN with an average black hole mass of $\sim 10^{10} \msol$ and bolometric luminosity of $\lbol \sim 10^{47.74} \lunit$. This sample therefore is vital for comparing the central engine properties of high redshift quasars having very massive black holes with that of the local LLQSOs.

\begin{figure*}
  \centering 
\includegraphics[width=9.0cm,angle=0]{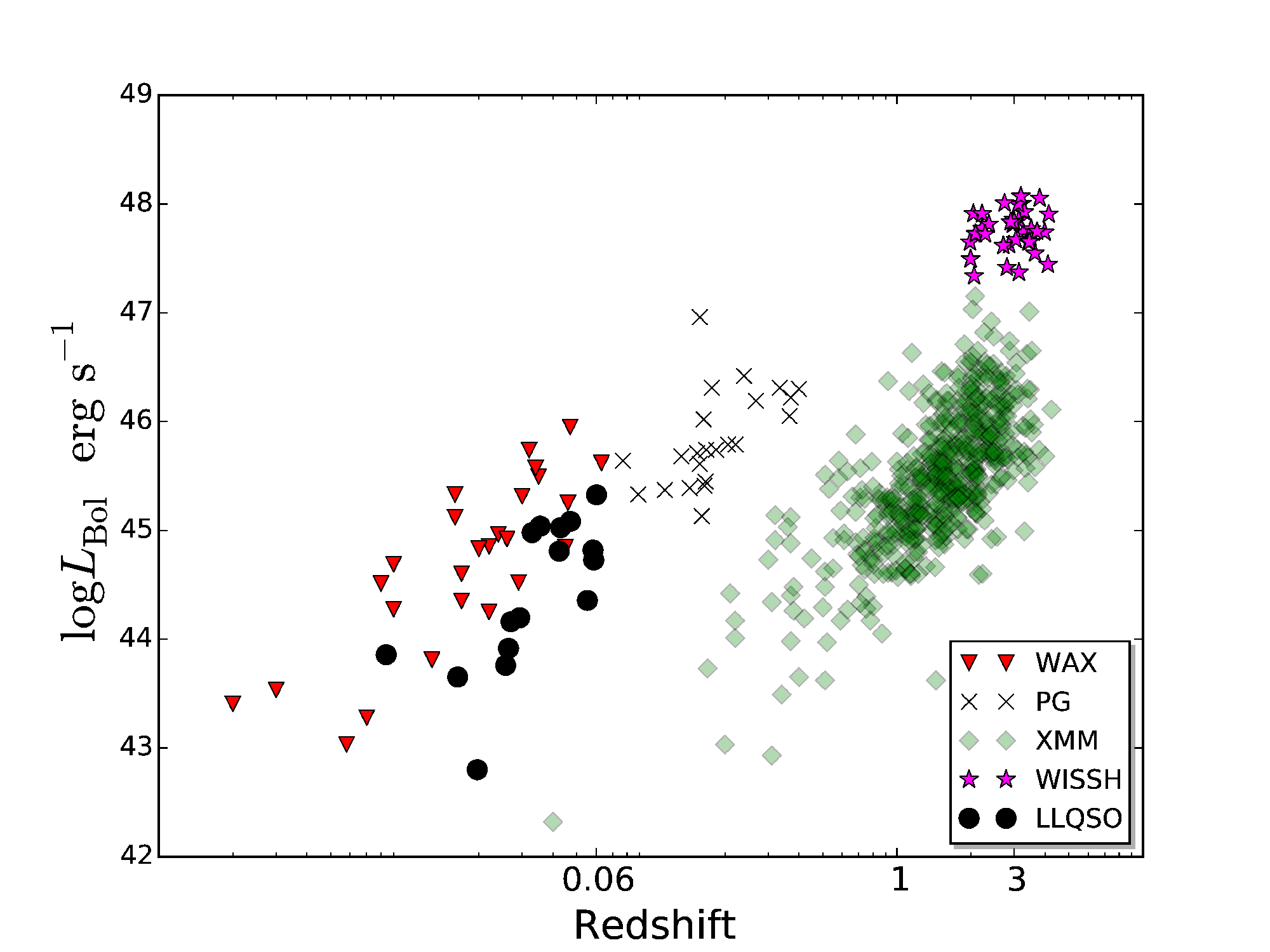}

	\caption{ The bolometric luminosity of the LLQSO and the other samples plotted against redshift. The samples are discussed in Section \ref{sec:sample}. The red inverted triangles denote WAX sources, the black circles denote the LLQSOs, the grey crosses denote the PG quasars, the green squares denote the XMM-COSMOS quasars and the magenta stars denote the WISSH quasars. We use these symbols consistently throught this work. Also see Tables \ref{Table:comparison} and \ref{Table:KS} and Section \ref{subsec:discussion1}. } \label{fig:evolution1}
\end{figure*}

As will be discussed below, Figure
  \ref{fig:evolution1} shows the bolometric luminosity $\lbol$ versus
  redshift diagram for the five samples LLQSO, WAX, PG-quasars,
  XMM-COSMOS and WISSH quasars. A 2 sample Kolmogorov-Smirnoff (KS) test showed that the LLQSOs have $\lbol$ different than the higher redshift
  bright quasars at a confidence $>99.99\%$, but similar to those of the nearby Seyfert galaxies.


\section{Observation and data reduction}\label{sec:obs}

 We list the observations of the sources in the LLQSO sample in Table
 \ref{Table:obs}. There are a total of 49 observations from the three
 telescopes \xmm{}, \chandra{} and \suzaku{}. We confine our study to observations obtained from 
 CCD and HXD detectors from these three telescopes to
 characterise the broadband X-ray spectral features, which is the main
 scientific aim of this work.

The EPIC-pn data from \xmm{} were reduced using the scientific
analysis system (SAS) software (version 15) with the task {\it
  epchain} and using the latest calibration database available at the
time we carried out the data reduction. We used EPIC-pn data because
of its higher signal to noise ratio as compared to MOS. We filtered
the EPIC-pn data for particle background counts using a rate cutoff of
$< 1 \, \rm ct s^{-1}$ for energy $>10\kev$, and created
time-averaged source + background and background spectra, as well as the
response matrix function (RMF) and auxiliary response function (ARF)
for each observation using the {\it xmmselect} command in SAS. The
source regions were selected with a circle radius of $40"$ centred on
the centroid of the source. The background regions were selected with
a circle of $40"$ located on the same CCD, but away from the
source. We checked for possible pile up in the LLQSO sources using the
command {\it epatplot} in SAS, and found that none of the sources are
piled up. The optical monitor (OM) camera simultaneously observed the sources along
with the EPIC camera. We reprocessed the OM data using the SAS task {\tt
  omichain} and used only fluxes measured by the UVM2 filter since its
  peak wavelength ($\rm 2300 \AA$) is nearest to $2500 \rm
  \AA$. For two sources, Mrk~1044 and Mrk~1239, where UVM2 filters were not used during the
  observations, we used fluxes obtained from the U and the V filters ($3440 \rm \AA$ and $5430 \rm \AA$ respectively), assuming a flat spectral slope in the UV-optical band.
  Host galaxy stellar contamination in the UV is likely minimal as
none of the LLQSO sources are known to host significant starburst activity. The
  observed UV fluxes were corrected for the Galactic reddening
  assuming \cite{1999PASP..111...63F} reddening law with $\rm R_v=
  3.1$ (see Table~\ref{Table:alphaox}).

The \chandra{} observations were reduced using the command {\it chandra\_repro} in the {\it CIAO} software (version 4.7.1) and using the latest calibration database. Source regions were selected using a circle of radius 2.5". The centroid of the circle was fixed to the RA and Dec of the source as obtained from NED. The background regions were selected using a circle of radius of 2.5" on the same CCD as the source, but away from the source. We used the command {\it specextract} to generate the source and background spectra along with the arf and rmf. We detected pile up in some of the \chandra{} spectra for the sources Mrk~1044 (observation id: 18685), Mrk~1018 (12868), NGC~0985 (12866)  and HE~1143$-$1810 (12873). We have used an annular source region for these cases with the inner radius of 0.5" and outer radius 2.5" to minimize the effect of pile up, which is predominant in the central pixels.

The \suzaku{} observations were performed using the X-ray Imaging Spectrometer (XIS) \citep{2007PASJ...59S..23K} and Hard X-ray Detector (HXD) \citep{2007PASJ...59S..35T}. The XIS observation were obtained in both the $3 \times 3$ and $5 \times 5$ data modes. The {\sc aepipeline} tool was used to reprocess and clean the unfiltered event files and to create the cleaned event files. In all observations, for both the XIS0 and XIS3(front-illuminated CCD) and for XIS1 (back-illuminated CCD), we extracted the source spectra for each observation from the filtered event lists using a $240$" circular region centered at the source position. We also extracted the corresponding background spectral data using four circular region of $120{\rm~arcsec}$ radii, excluding the source region. There is no pile up in \suzaku{} observation for the sources in the LLQSO sample.

The \xmm{} spectra were grouped by a minimum of 20 counts per channel and a maximum of five resolution elements using the command {\it Specgroup} in SAS. The \chandra{} and \suzaku{} spectra were grouped by a minimum of 20 counts per channel in the ISIS (Interactive Spectral Interpretation System) software \citep{2000ASPC..216..591H}.


\section{X-ray data analysis}\label{sec:data}

We used a set of phenomenological and physical models to describe the continuum as well the discrete components in the X-ray spectra of the LLQSOs. The baseline model consists of a neutral absorption due to the Galaxy ({\it tbabs}), a neutral absorption intrinsic to the galaxy ({\it ztbabs}), the soft X-ray excess described using a black body ({\it bbody}), the coronal emission described using a power law ({\it power law}). The blackbody model is known to be unphysical, however it is sufficient for our purpose to describe the soft X-ray spectra. The ionized absorption features when detected, were modeled using warm absorber table models developed using CLOUDY. For sources detected with Compton hump due to scattering of hard X-ray photons off a neutral medium, we modeled the narrow FeK$\alpha$ emission line and the Compton hump self consistently with the model {\it MYTorus} \citep{2009MNRAS.397.1549M, 2012MNRAS.423.3360Y}. However, when Compton hump was not detected, we modeled the FeK $\alpha$ emission lines using Gaussian profiles.  The scattered and Fe K emission line components ({\it MYTorusS} and {\it MYTorusL}) describe the neutral reflection from distant matter lying out of the line of sight. The {\it MYTorus} inclination angle and the normalisation of the individual components were left free to vary. The incident power law slope $\Gamma$ in {\it MYTorus} was tied with the primary power law component. However, 
as a caveat we should note that the
reflection fraction of the distant neutral reflector is best
constrained by X-ray continuum emission beyond $10\kev$, which in the
present situation can be acheived by HXD-PIN data from \suzaku{}
telescope. Only six sources out of 16 have been observed by
\suzaku{} (see Table \ref{Table:obs}). For the remaining sources, which lack
$>$10 keV data,  failure to model
a hard excess may lead to a slight artifical flattening of $\Gamma$ by typically $\sim0.1-0.2$.
We also note that only 8 out of
17 sources in the LLQSO sample have publicly available data from the
\nustar{} observatory, which covers an energy range of $\sim
3-80\kev$, ideal for constraining the hard X-ray excess due to Compton
reflection. However, as \nustar{} does not give us a simultaneous spectral view in
the soft energy range $\le 3 \kev$, which is necessary to study the soft X-ray properties of the sources and to detect the presence of any neutral absorption intrinsic to the host galaxy. Therefore, we have not used any observations
from this telescope. 

The warm absorber table model used in this work was created using the modeling code CLOUDY \citep{1998PASP..110..761F,2017arXiv170510877F}. The input spectral energy distribution in the energy range $1-1000$ Ryd was that of a typical Seyfert galaxy, Mrk~704 \citep{2011ApJ...734...75L}. We know that the exact characterisation of ionisation parameter $\xi$ and column density $\nh$ of the warm absorbers are dependent on the shape of the SED \citep{2013ApJ...777....2L}. However, the systematic uncertainties introduced by this approximation are acceptable because we are using only CCD-resolution X-ray data in this paper. 

All errors quoted on the fitted parameters reflect the $90\%$ confidence interval for one interesting parameter, corresponding to $\Delta \chi^2=2.7$ \citep{1976ApJ...208..177L}. The ISIS software \citep{2000ASPC..216..591H} was used in fitting the spectra. Table \ref{Table:alphaox} lists the $2500\rm \AA$ and $2\kev$ monochromatic fluxes along with the $\alpha_{\rm OX}$ values of the LLQSOs. Tables \ref{Table:xray1}, \ref{Table:xray2} and \ref{Table:xray3} list the results from the X-ray spectral analysis and the X-ray and bolometric luminosities of the LLQSOs. Table \ref{Table:mol} lists the mass of the molecular gas present in the LLQSOs.


\section{Results}\label{sec:results}
\begin{figure*}
  \centering 
	\hbox{
\includegraphics[width=9.0cm,angle=0]{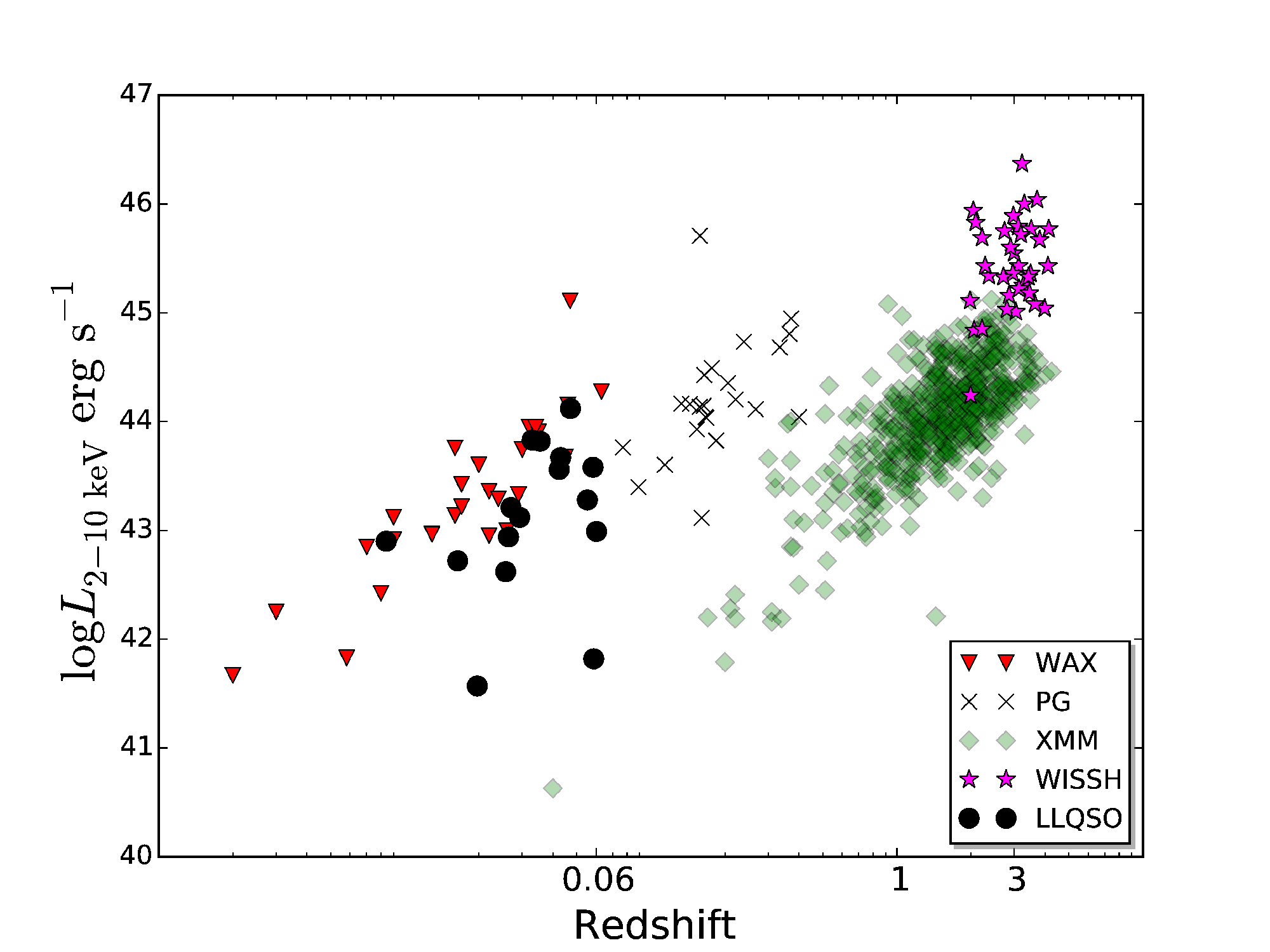} 
\includegraphics[width=9.0cm,angle=0]{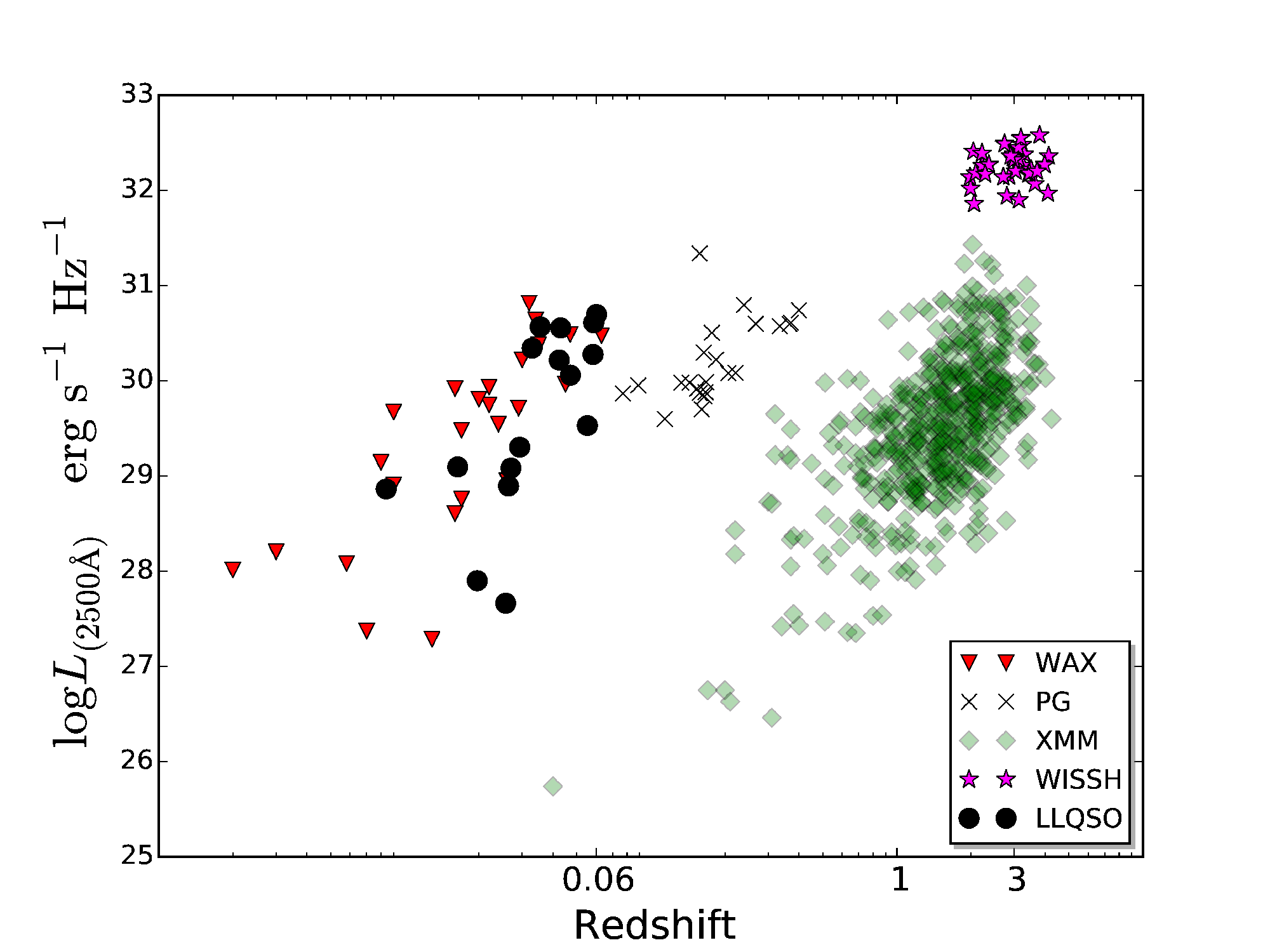} 
}
	\caption{{\it Left:} The $2-10\kev$ luminosity of the LLQSO and the other samples plotted against redshift. Also see Tables \ref{Table:comparison} and \ref{Table:KS} and Section \ref{subsec:discussion1}. {\it Right:} Same as left, except for the Y axis, which is $2500 \,\rm \AA$ luminosity.} \label{fig:evolution2}
\end{figure*}

We have carried out a broadband X-ray spectral analysis of the LLQSO sources with all the available observations from the three telescopes \xmm{}, \chandra{} and \suzaku{}. Figures \ref{fig:evolution1} and \ref{fig:evolution2} show the distribution of various X-ray and  UV parameters of the LLQSO sources and a comparison with the other samples, as described in Section \ref{subsec:controlsamples}. 

We have used the freely available Python code by \citet{2012Sci...338.1445N} using the BCES technique \citep{1996ApJ...470..706A} to carry out the linear regression analysis between several source parameters. In this method the errors in both variables defining a data point are taken into account, as is any intrinsic scatter that may be present in the data, in addition to the scatter produced by the random variables. The strength of the correlation analysis was tested using the non-parametric Spearman rank correlation method. We declare a correlation to be significant if the null hypothesis probability is rejected at a confidence greater than $99\%$. Below we discuss some of the main results of the LLQSO sample study.

\subsection{The X-ray properties of the LLQSOs}
We first discuss the results of the X-ray spectral fits.
The X-ray power-law slope distribution of the LLQSO sources ranges from
$\Gamma=1.45 -2.4$, with a mean value of 1.74 and a standard deviation
of 0.32. 

After correcting for the Galactic extinction, we found that
only three sources required an additional intrinsic neutral absorption
column (See Table \ref{Table:xray1}). These sources, NGC~0985,
Mrk~1239 and HE~1136--2304 required best-fit column densities of
$1.19_{-0.11}^{+0.12}\times 10^{21}\cmsqi$,
$3.77_{-0.09}^{+0.11}\times 10^{21} \cmsqi$ and
$1.22_{-0.09}^{+0.13}\times 10^{21}\cmsqi$, respectively. For six
sources in the LLQSO sample we detected ionised absorption in the soft
X-rays. In two of these cases two components were required to fit the ionized absorption, while only one component was needed in the other four cases. (see Table
\ref{Table:xray2}). The detected warm absorbers span ionisation parameters
$\log\xi \sim 0.79-3.38 \xiunit$ and column 
densities $\log\nh \sim 20.62-22.65 \cmsqi$.


We detect neutral narrow Fe~K emission lines at $\sim 6.4\kev$ in all but two
of the LLQSOs (Mrk~618 and PG~1011--040). In four sources, we detect high ionisation Fe
K$\alpha$ emission lines and neutral Fe K$\beta$ emission lines. For two
sources we detect broad Fe K emission lines which were modeled using
{\it diskline} profile \citep{1989MNRAS.238..729F}. We find the
presence of a hard X-ray excess due to the reflection of primary X-ray
photons off a distant neutral reflector in five of our sources, though
the reader is directed to the caveats above regarding the lack of $>$10 keV data. See Tables \ref{Table:xray1} and \ref{Table:xray2} for
details


There are 11 sources for which there are two or more observations. 
As can be seen from Table \ref{Table:xray2},
these LLQSO sources are overall not significantly variable in X-ray flux between
different observations, with a few exceptions.
The maximum inter-observation variability in the soft
X-rays is recorded for the sources NGC~0985 ($\sim 77\%$) and
HE~1143--1810 ($\sim 89\%$). For the same sources, the $\lhard$
luminosity has not varied more than $\sim 34\%$.

We note that two sources in the LLQSO sample, Mrk~1018
and HE~1136--2304, have been characterized
as changing look AGN by previous studies. However, previous studies were not able to
definitively ascertain whether variable obsuration or variations in instrinsic luminosity were
ultimately responsible for the drastic observed changes:
\citet{2016A&A...593L...8M}
found that Mrk~1018 had returned to the optical spectral classification of Seyfert 1.9 in 2015 after
almost 30 years. The broad and narrow optical emission lines which
were detected in the source spectrum earlier had completely
disappeared in a recent observation in 2015. The most recent
observation of Mrk~1018 by \chandra{} in 2016 also shows no source
photons in X-rays. \citet{2016MNRAS.461.1927P} found from the 
long term light curve of HE~1136--2304 that between 1993 and 2015, the
source changed its classification from Seyfert 2 to Seyfert 1.5, with
emergence of broad Balmer lines in the recent observations. The recent X-ray
observations of this source with \xmm{} and \nustar{} indicate the
presence of a moderate neutral obscurer in the X-rays, with a column
density of $\nh \sim 10^{21} \cmsqi$. 

 The soft X-ray ($0.3-2\kev$) luminosity of the LLQSOs ranges from $\log\lsoft=42.0-44.4 \lunit$, while the hard
X-ray ($2-10\kev$) luminosity ranges from $\log\lhard=41.5-44.4 \lunit$. Table
\ref{Table:xray3} lists the soft X-ray ($\lsoft$), hard X-ray
($\lhard$), and bolometric luminosities ($\lbol$) of the LLQSO
sources. The bolometric luminosities are estimated using the relation
$\lbol= \kappa_{\rm Lbol} \times \lhard$, where $\kappa_{\rm Lbol}$ is
the bolometric correction factor. The value of $\kappa_{\rm Lbol}$
for each source is obtained from the scaling relation

\begin{equation}\label{equ:kappa}
\log \kappa_{\rm Lbol}=1.561-1.853\times \alpha_{\rm OX} + 1.226 \times \alpha_{\rm OX}^2,
\end{equation}

\noindent \citep{2010A&A...512A..34L} where $\alpha_{\rm OX}$ is the power-law slope joining the $2\kev$ and the $2500\rm \AA$ flux for a given source (See section \ref{subsec:alphaox} for details).

We also calculate $\lambda_{\rm Edd}=\lbol/\ledd$; values are listed in Table~5. 
We find that the LLQSO sources are mostly sub-Eddington, with $\lambda_{\rm Edd} \sim 0.003-0.389$. One source, HE~1143--1810, shows
super Eddington rates, $\sim 1.46$. Three sources in the sample (Mrk~1044, Mrk~1298, Mrk~0926) show near-Eddington accretion rates.

\begin{figure*}
  \centering 

\includegraphics[width=10.0cm,angle=0]{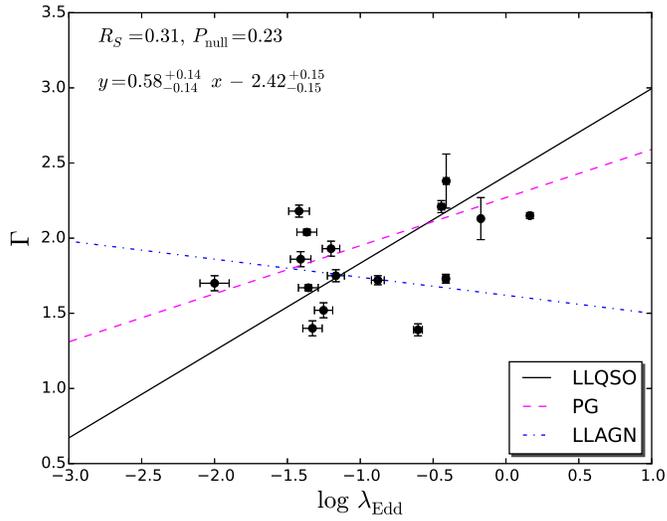} 

	\caption{ {The logarithm of the Eddington ratio vs the power law slope of the LLQSO sources. The black solid line is the best fit correlation slope for the LLQSO sources. The pink dashed line is the emperical relation obtained from \citet{2013MNRAS.433.2485B}, while the blue dash-dotted line with a negative slope is the relation obtained for low luminosity AGNs, LLAGNs \citep{2009MNRAS.399..349G}. Note that the LLQSOs lie mostly at the inflection point of positive correlation (for Seyferts) and negative correlations (for LLAGNs). See Section \ref{sec:discussion} for details.}} \label{fig:GammaLambdaEdd}
\end{figure*}

\begin{figure*}
  \centering 

\includegraphics[width=10.0cm,angle=0]{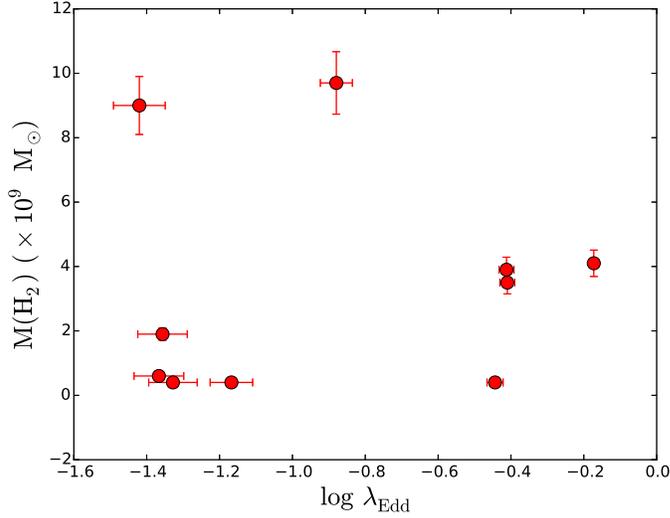}

	\caption{ The Eddington ratio vs the molecular gas mass M(H2) in the host galaxy of the LLQSOs. The errors on M(H2) are assumed to be $10\%$ on the values quoted in Tables \ref{Table:mol}. We do not detect any significant correlation between the quantities. } \label{fig:Molgasmass}
\end{figure*}

Figure \ref{fig:GammaLambdaEdd} shows $\Gamma$ plotted against $\log \lambda_{\rm Edd}$. The best-fit correlation slopes, intercepts and the non-parametric Spearman correlation strengths are quoted in the figure. We find that the null hypothesis probability cannot be rejected at sufficient confidence level, implying that the correlation is not statistically robust. We do not detect any correlation between $\lambda_{\rm Edd}$ and molecular gas ($M(H2)$) present in the host galaxy (See Figure \ref{fig:Molgasmass}). We discuss the possible reasons in the Discussion section. In all cases of correlations, we have assumed a $5\%$ error on $\lambdaedd$ and a $10\%$ error on $M(H2)$.

\subsection{The UV flux and $\alpha_{\rm OX}$ distribution}\label{subsec:alphaox}

Table \ref{Table:alphaox} lists the UV monochromatic fluxes at $2500
\rm \AA$ for the longest \xmm{} observations of the LLQSO sources.  The absorption-corrected UV and 2 keV fluxes
were used to calculate $\rm \alpha_{\rm
  OX}=-0.384 \log[ L_{2500 \rm \AA}/L_{2\kev}]$
\citep{1979ApJ...234L...9T}. Figure \ref{fig:alphaoxcorr} shows the correlation between
$\alpha_{\rm OX}$ vs $\rm L_{2500\rm \AA}$ for the LLQSO sources. The
best-fit linear regression slope and intercept is given by:
$\alpha_{\rm OX}=-0.29_{-0.08}^{+0.08}\log(L_{2500\rm
  \AA})+7.36_{-2.37}^{+2.37}$. The correlation is strong, with a null
hypothesis probability of $P_{\rm null}\sim 3\times 10^{-4}$. We compare our results on $\alpha_{\rm OX}$ to other AGN samples
in the Discussion section below.


\section{Discussion}\label{sec:discussion}

From Fig \ref{fig:evolution1} we find that the LLQSOs are local quasars with lower $\lbol$ compared to the higher redshift quasars and hence can potentially shed light on how differently the central engines of the quasars function with redshift \citep{2012nsgq.confE..69M,2014A&A...561A.140B,2015A&A...580A.113T,2016A&A...587A.138B}. Below we discuss the possible reasons for the relative weakness in luminosity of the LLQSOs in the light of their accretion disk and corona properties. We also discuss the effects of the presence or absence of molecular gas in the host galaxy on the accretion states of the black holes.

\subsection{The nature of the central engines of the LLQSOs}\label{subsec:discussion1}




In this Section we explore the nature of the central engines of the LLQSOs by comparing the distributions of physical quantities such as, bolometric luminosity $\lbol$, black hole mass $\mbh$, etc., which are diagnostics of accretion efficiency. The average values and the $1\sigma$ dispersion of these parameters for all the samples are listed in Table \ref{Table:comparison}. Results of KS tests comparing distributions of various parameters from the LLQSO sample against those from other samples are listed in Table \ref{Table:KS}. We also investigate the correlations between the various parameters, as discussed below.

\subsubsection{The distributions of $\lhard$, $\lbol$, $\luv$, $\mbh$, $\alpha_{\rm OX}$ and $\lambdaedd$} \label{subsubsection:distributions}

Figures \ref{fig:evolution1} and \ref{fig:evolution2} show evolution of the $\lbol$, $\lhard$ and $\luv$ with redshift for the LLQSO and the PG, XMM-COSMOS and WISSH quasar samples. Table \ref{Table:KS} shows that the $\lbol$ and $\lhard$ of the LLQSO sources originate from a different parent sample as that of the PG, the XMM-COSMOS and the WISSH quasars, with the KS test null hypothesis being rejected at $>99.99\%$ confidence. We also find that the $\lbol$ and $\lhard$ luminosities of the LLQSOs have similar parent population as that of the local Seyfert galaxies (WAX). From Table \ref{Table:comparison} we find that for
the PG, XMM-COSMOS, and WISSH quasar samples the average and $1\,\sigma$ dispersion in hard X-ray
luminosity $\log\lhard$ are $44.22\pm0.53$, ${44.04\pm0.53}$, and
$45.44\pm0.41\lunit$, respectively. On the other hand, $\log\lhard$ for
the LLQSOs is $43.10\pm0.68\lunit$, similar to that for
WAX, but nearly one order of magnitude lower than the PG and XMM-COSMOS
quasars, and two orders of magnitude weaker than the WISSH
sample.

One possiblity for this relative weakness in luminosity compared to the
bright quasars could be that LLQSOs are obscured, but this can be
ruled out given the fact that we did not detect any intrinsic neutral
absorption column density in the X-rays for any source greater than
$10^{22} \cmsqi$. Therefore we are looking directly at the central engines of
these sources.

However, there are a few potential exceptions. The
  sources Mrk~1018 and HE1136-2304 have been recently classified as
  changing look sources, with the previous studies not being able to
  finally dissect whether it is the changing obscuration or the
  changing luminosity of the source that is responsible
  \citep{2016A&A...593L...8M,2016MNRAS.461.1927P}. The source ESO~113-G010 in
  the LLQSO sample is another interesting source, as it is classified
  as Seyfert 1.8 from optical observations. 
 \citet{2012A&A...542A..30M} found a large
 Balmer decrement ($H\alpha/H\beta \sim 8$) in this galaxy, indicating
 a significant amount of absorption along the line of sight. They measured
only ionized absorbing components of columns $\sim10^{22}$ cm$^{-2}$, with no neutral absorption (similar to our 
analysis), concluding that a dusty warm absorber is responsible.

Most interestingly from Table \ref{Table:KS} we find that the distribution of $\log\lambdaedd$ of the LLQSOs are similar to that of the WAX, the PG as well as the XMM-COSMOS samples, indicating that the accretion rates at the heart of the central engine of these powerful quasars at different redshifts are similar. If the values of $\lambdaedd$ of these sources are similar, then possibly the LLQSOs are scaled down version of the more massive higher redshift quasars, which is also corroborated by Table \ref{Table:KS} showing that the distribution of the $\log\mbh$ of the higher redshift quasars are different (higher) from that of the LLQSOs. The average values and $1\, \sigma$ dispersion of the black hole masses $\log(\mbh/\msol)$ obtained for the different samples are : LLQSOs $7.35\pm0.65$, PG $8.32\pm0.53$, XMM-COSMOS $8.41\pm 0.39$ and the WISSH quasars $9.98\pm0.43$ (See Table \ref{Table:comparison}).

\subsubsection{The $\alpha_{\rm OX}$-$L_{2500 \rm \AA}$ anticorrelation}\label{subsubsec:alphaox}

\begin{figure*}
  \centering 

\includegraphics[width=10.0cm,angle=0]{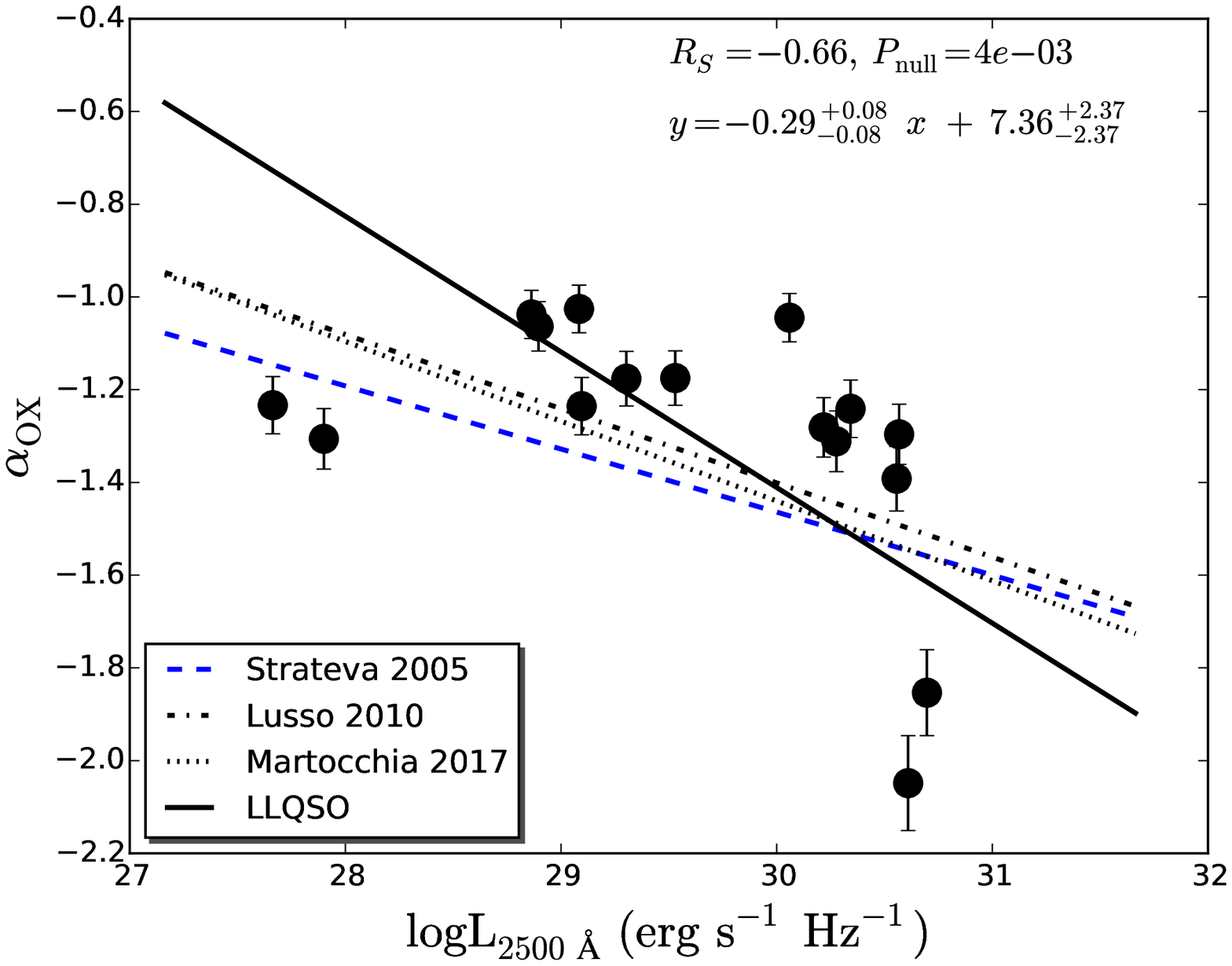} 

	\caption{ The anticorrelation between the $2500 \, \rm \AA$ luminosity and $\alpha_{\rm OX}$ of the LLQSOs. The solid line is the best fit correlation slope for the LLQSO sources only. The dashed line, dash-dotted line and the dotted line are the best fit correlation slopes obtained from \citet{2005AJ....130..387S}, \citet{2010A&A...512A..34L} and \citet{2017A&A...608A..51M} respectively for different AGN samples.  See Section \ref{sec:discussion} for details. Note that the errors on $\luv$ are smaller than the circle size and hence not visible.} \label{fig:alphaoxcorr}
\end{figure*}

From a sample of AGN over a redshift spread of $z=0-6.2$,
\citet{2005AJ....130..387S} found that $\alpha_{\rm OX}$ decreases
with the UV monochromatic luminosity $L_{2500 \rm \AA}$. The
$\alpha_{\rm OX}$-$L_{2500 \rm \AA}$ anticorrelation implies that the
relative proportions of UV and X-ray emission depend on bolometric
luminosity, as has also been found by previous studies of AGN SEDs \citep[See for e.g.,][]{2004MNRAS.351..169M,2016ApJ...819..154L,2017A&A...602A..79L}. Relatively more luminous AGN will emit relatively fewer X-rays. Studies such as \citet{2005AJ....130..387S} and \citet{2016ApJ...819..154L} also found that the slope of the $\alpha_{\rm OX}$-$L_{2500 \rm \AA}$
anticorrelation does not depend on the average redshift of the AGN
sample on which it is calculated, implying that the central engine of
AGN functions similarly through out the cosmic time and hence is a good
benchmark to test whether a source is X-ray weak or strong relative
to the UV flux. 
Figure \ref{fig:alphaoxcorr} shows the correlation between $\alpha_{\rm OX}$ and $L_{2500 \rm \AA}$ for the LLQSO sources with the best fit linear regression line in black line. We also plotted the best fit linear regression slope from three other quasar samples, the XMM-COSMOS quasars \citep{2010A&A...512A..34L}, the WISSH quasars \citep{2017A&A...608A..51M} and the optically selected quasars \citep{2005AJ....130..387S}, which are similar to each other within their statistical errors. The best fit linear regression slope obtained by \citet{2005AJ....130..387S} is $\alpha_{\rm OX}=-0.136_{-0.013}^{+0.013}\log L_{2500\rm\AA}+2.616_{-0.398}^{+0.398}$. We note that the best-fit anti-correlation for the LLQSOs is $\alpha_{\rm OX}=-0.29_{-0.08}^{+0.08} \log L_{2500 \rm \AA}+7.36_{-2.37}^{+2.37}$ detected at a confidence $>99.99\%$. The strong anticorrelation leads us to conclude that the central engine of the LLQSOs function similar to that of other quasars. From Figure \ref{fig:alphaoxcorr} we note two things: A. The correlation slope of the LLQSOs is slightly steeper than those of the other quasar samples, and B. 12 out of 16 LLQSO sources lie above the best-fit linear regression derived by \citet{2005AJ....130..387S}, and 11 out of 16 sources lie above all the three correlations slopes derived for other quasar samples. These results are interesting as they possibly indicate that most LLQSOs are efficient X-ray emitters for a given UV luminosity, when compared to other quasars. Speculatively, this could be due to a more efficient coupling between the disk UV photons and the X-ray corona.

\subsubsection{The $\lbol$ vs X-ray bolometric corrections}

\begin{figure*}
  \centering 

\includegraphics[width=10.0cm,angle=0]{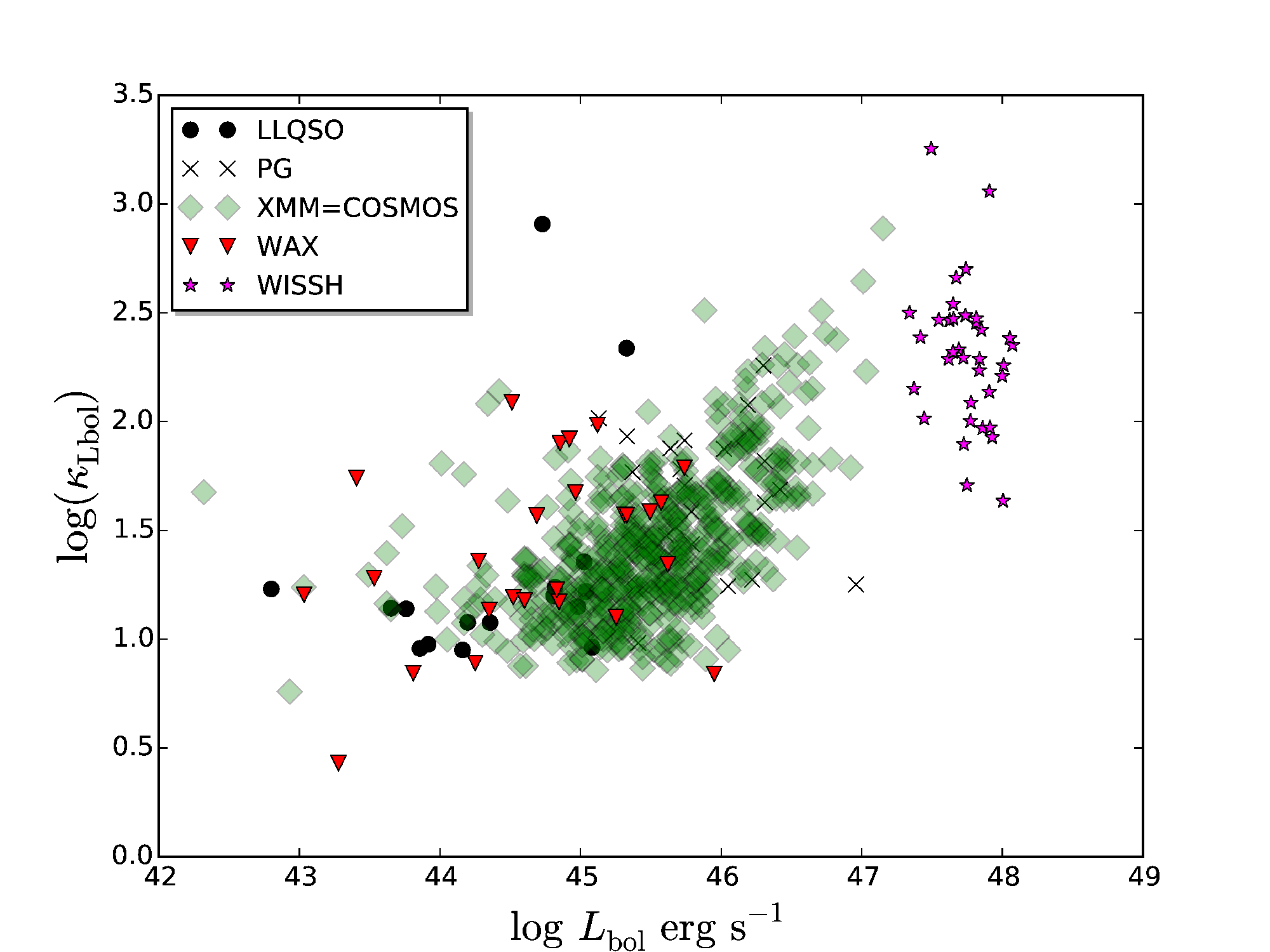}

	\caption{The relationship between the hard X-ray bolometric correction factor ($\kappa_{\rm Lbol}$) with the bolometric luminosity $\lbol$. }    \label{fig:kappalbol}
\end{figure*}

Several quasar studies have found that with increasing bolometric luminosity, the  bolometric correction $\kappa_{\rm Lbol}$ increases \citep{2017A&A...608A..51M,2012MNRAS.425..623L}. This would imply that with increasing $\lbol$, the corona radiative power represented by $\lhard$ becomes weaker relative to the optical-UV disk emission and hence a larger correction factor becomes necessary. The physical interpretation of this relation is similar to the $\alpha_{\rm OX}$-$L_{2500 \rm \AA}$ anticorrelation derived in Section \ref{subsubsec:alphaox} where we found that a stronger UV emission will lead to a weaker X-ray luminosity. From Figure \ref{fig:kappalbol} we find that the LLQSOs follow a similar trend in the $\lbol$ vs $\kappa_{\rm Lbol}$ relation, indicating that the central engines of these quasars function similar to those at higher redshift. However, as a caveat we must note that the bolometric luminosities of the LLQSOs have been derived using the relation $\lbol= \kappa_{\rm Lbol} \times \lhard$, where $\kappa_{\rm Lbol}$ has been estimated using equation \ref{equ:kappa}, hence there is an intrinsic dependence of $\lbol$ on $\kappa_{\rm Lbol}$.

\subsubsection{$\Gamma$ vs the Eddington ratio $\lambda_{\rm Edd}$}
   
The relation between the $2-10\kev$ power law slope, $\Gamma$, and the
Eddington ratio ($\lambda_{\rm Edd}$) provides another test to check
how efficiently the disk photons are coupled with the hot corona and hence how efficient the central engines are. A strong coupling between $\Gamma$ and $\lambda_{\rm Edd}$,
implies that a higher accretion rate cools off the corona faster,
leading to steeper power-law slopes \citep{1995MNRAS.277L...5P}.

\citet{2013MNRAS.433.2485B} have studied
the relation between $\Gamma$ and $\lambda_{\rm Edd}$ in a sample of
radio quiet AGN up to redshift $z\sim 2$. They found a strong correlation
quantified as $\Gamma=(0.32\pm0.05)\log\lambda_{\rm Edd} + (2.27\pm 0.06)$. In a recent study, of a BAT selected AGN sample, \citet{2017MNRAS.470..800T} have found similar strong correlations. From Figure \ref{fig:GammaLambdaEdd}, however, we find that the LLQSOs do not show
any strong correlation between the spectral slope and the Eddington rate. It is possible that our sample
suffers from small-number statistics and/or an insufficient range in 
$\lambda_{\rm Edd}$.

\citet{2009MNRAS.399..349G} investigated the $\Gamma$-$\lambda_{\rm
  Edd}$ for a sample of 57 low luminosity AGN (LLAGN) in the local
Universe and found that they follow an anti-correlation. This is in
contrast to the positive correlation obtained for bright AGN,
suggesting two modes of accretion above/below some critical transition
value of $\lambda_{\rm Edd}$, likely $\sim0.01-0.1$; LLAGN emission
could be dominated by processes associated with advection dominated
accretion flows (ADAF).  We note that the LLQSOs span $\lambda_{\rm
  Edd} \sim 10^{-2}-1$ and $\Gamma \sim 1.20-2.04$, and roughly overlap
with the inflection point in the $\Gamma$-$\lambda_{\rm Edd}$ relation
of \citet{2009MNRAS.399..349G}; LLQSOs' having $\lambda_{\rm Edd}$
near the critical transition value could 1) explain why we do not see
any strong correlation or anti-correlation, and 2) indicate that it is
unclear what type of accretion flow and emission process dominate,
similar to objects near the Seyfert/LLAGN boundary in the local
Universe. A larger sample of LLQSOs spanning a much larger range in 
$\lambda_{\rm Edd}$ can potentially yield more insight. \\

In summary we find that: 1. The LLQSOs are less luminous compared to their higher redshift counterparts mostly because of their smaller black hole size. They may be the scaled down versions of the higher redshift quasars, 2. The central engines of the LLQSOs function similarly as that of the higher redshift quasars, as evident from the similar distribution of $\lambdaedd$, and the relations between $\alpha_{\rm OX}-\luv$ and $\lbol-\kappa_{\rm Lbol}$. However, the hyperluminous WISSH quasars probably have more efficient central engines with higher accretion rates, 3. The nature of accretion of the LLQSO is uncertain as we do not find any trend in the $\Gamma-\lambdaedd$ relation, 4. Possibly the corona of the LLQSOs are efficiently coupled with the disk photons. 


\subsection{The link between the accretion states and presence of molecular gas.}\label{subsec:moleculargas}

The presence or absence of
molecular gas may play a leading role in defining the accretion
state of the black hole.
11 out of 17 sources in the LLQSO sample have been observed in the IR
by \citet{2007A&A...470..571B} and have been detected with the
presence of large amounts of molecular gas in the host galaxy. The
authors have studied the $\rm ^{12}CO(1-0)$ and $\rm ^{12}CO(2-1)$ molecular
emission lines for these sources using the IRAM 30 metre telescope. Table \ref{Table:mol} shows
  the detected molecular gas mass, and it ranges over an order of
  magnitude, $(0.4-9.7)\times 10^9 \msol$. However, how the $\kpc$ scale gas
  efficiently loses angular momentum and flows into the SMBH accretion
  disk at $<\pc$ scale is still a matter of debate. Thus the sheer
  presence of molecular gas may not mean efficient accretion unless
we observe a direct relation between the presence of molecular gas
and the rate of accretion onto the SMBH.

A sample study of the molecular gas of high redshift quasars ($z\sim
1.5$) were carried out by \citet{2017MNRAS.468.4205K} using ALMA
observations of 10 sources. The redshift range of $z=1-2$ is believed
to be the quasar peak era when the occurrence fraction of bright AGN
as well as the average accretion rate of the AGN were higher than that
of the present epoch. Those authors found that the galaxies that
host AGN have a lower molecular gas fraction, implying that AGN feedback may
have depleted the gas reservoir in the host galaxy.
Regarding our LLQSO sample, in Figure \ref{fig:Molgasmass}, we tested
for any anticorrelation between $\lambda_{\rm Edd}$ and the molecular
gas mass which could support the study of
\citet{2017MNRAS.468.4205K}. However, we do not detect any such anti-correlation. 

We must note that we only have access to current ``snapshots'' of $\lambda_{\rm Edd}$ and thus we are using it
as a proxy for a long-term average value of accretion rate, despite the fact that AGN luminosity is known to ``flicker''
on timescales of $\sim 10^5$ years \citep[][]{2015MNRAS.451.2517S}, and we do not have a solid handle on timescales over which
molecular gas can be transported from large-scale reservoirs down to the SMBH and/or blown out by AGN feedback.


Most of the sources in the LLQSO sample have significant amounts of 
molecular gas, $(0.4-9.7)\times 10^9 \msol$, along with
moderate accretion rates. We now raise the possibility that
the weakening of the AGN phase (compared to their higher redshift counterparts) is not caused simply by the total absence
of molecular gas, but instead due to the
absence of an effective physical mechanism by which
the gas can infall from the large scale host galaxy down to the SMBH's accretion disk.
We consider the case of the LLQSO Mrk~590,
for which we measure a relatively low value of $\lambda_{\rm Edd}$, $\sim 6\times 10^{-2}$,
and whose host galaxy is known to host a molecular gas mass of
$1.9\times 10^9\msol$ \citep{2007A&A...470..571B}. 
Mrk~590 is also
classified as a changing look AGN: from 2006 to 
2012, the broad $H\beta$ emission line in the optical spectrum has
disappeared \citep[See][ and the references
  therein]{2014ApJ...796..134D}. The reason for its variable nature in
optical, as argued by \citet{2014ApJ...796..134D} is more likely due
to changes in the state of accretion due to the absence of fuel,
rather than varying obscuration by clumpy gas clouds. The absence of
fuel can occur either if the black hole has used up all the cold gas
in its vicinity, or if the cold gas has been blown off by energetic
outflowing winds. \citet{2016MNRAS.455.2745K} studied the central
$\sim 500\pc$ region of Mrk~590 in radio with the ALMA observatory to
investigate the presence of cold gas in the vicinity of the black hole
and its effect on the accretion state. They could constrain a
molecular gas mass of $\le 1.6\times 10^5 \msol$ in the inner $150
\pc$, which they conclude to be potentially enough to feed the central
SMBH for another $2.6\times 10^5$ years assuming Eddington limited
accretion. However, $\lambda_{\rm Edd}$ is low, comparable to local
Seyfert galaxies, which leads them to conclude that Mrk~590 is
going through a temporary feeding break, and
that perhaps the gas at $150\pc$ does not have any impact on the SMBH
accretion rate, which is controlled by sub-pc mechanisms.

In summary, 
1) we find no direct evidence in the LLQSO sample
that AGN accretion rate relative to Eddington is linked to the molecular gas mass, and 
2) even though most LLQSOs have massive
reservoirs of molecular gas in their host galaxies, they may not possess an
efficient mechanism to accrete them onto the super massive black
hole.


\section{Conclusions}\label{sec:conclusion}

We have carried out a systematic X-ray spectral analysis of the LLQSO
sample of sources and investigated why these sources are weaker in terms of overall luminosity
compared to the high redshift quasars in view of their accretion state
and disk-corona properties. The LLQSOs are a subsample of the
Hamburg/ESO survey (HES) for bright quasars, with a redshift cut-off
$z<0.06$ and consist of 99 AGN. The present work deals with 16 of
these sources that have publicly available data in the archives of
\xmm{}, \chandra{} and \suzaku{}. We list below the main conclusions
of this paper:

\begin{itemize}

	\item The $\lhard$, and $\lbol$ luminosities of the LLQSOs are lower compared to the higher redshift quasars by almost an order of magnitude, but are similar to the local Seyfert galaxies.

	\item The distribution of the Eddington rate $\lambdaedd$ of the LLQSOs is similar to that of the local Seyfert galaxies and also the higher redshift quasar samples PG and XMM-COSMOS with a KS test confidence of $>99\%$. The central engines of the LLQSOs therefore possibly function similarly as that of the higher redshift quasars, and the reason for lower luminosity could probably be due to their lower black hole mass. Speculatively the LLQSOs may be regarded as scaled down versions of the higher redshift quasars.

	\item The best fit $\alpha_{\rm OX}$ vs $L_{2500 \rm \AA}$ anti-correlation
  of the LLQSO sources is $\alpha_{\rm OX}=-0.29_{-0.08}^{+0.08}
		L_{2500 \rm \AA}+7.36_{-2.37}^{+2.37}$, detected with a confidence of $>99.99\%$. The anti-correlation between these quantities indicates that as the UV luminosity of the source increases, the X-ray luminosity decreases, which has also been detected in several other quasar studies with large sample sizes. However, the slope obtained with LLQSOs is slightly steeper (within $2\, \sigma$) than the other quasar samples, and 12 out of 16 sources in the LLQSO samples lie above the linear regression line obtained for other quasar samples. This may possibly indicate that with respect to the other quasar samples, the LLQSOs are comparably efficient X-ray emitters for a given UV luminosity. This may be explained by a better coupling between the disk photons and the X-ray-emitting corona. This will be tested in future with a larger LLQSO sample.

\item The X-ray power law spectral index, $\Gamma$, and the Eddington
  rate for the LLQSOs do not show any strong correlation. 
The lack of a correlation could be due to the small number in the sample, and/or because
the values of $\lambda_{\rm Edd}$ span the
critical transition value between different modes of accretion
posited for AGN in the local Universe (e.g., ADAFs versus radiatively-efficient disks in LLAGN and Seyferts, respectively.)

   \item The LLQSOs are mostly unobscured in X-rays in terms of neutral obscuration, with three exceptions:
NGC~0985, Mrk~1239 and HE~1136--2304, which have obscuring column
 densities on the order of $\sim 10^{21}\cmsqi$. However, two of the
  LLQSOs, Mrk~1018 and HE~1136--2304 are changing look in nature, and
  previous multi-wavelength studies of these sources could not distinguish between
		changes in the accretion states of the AGN versus obscuration.

\item Warm absorbers are not ubiquitous in these sources. Only 5 out
  of 16 LLQSOs exhibit signatures of ionised absorption in soft
		X-rays.

\item The presence of molecular gas in the host galaxy does not
  significantly influence the (instantaneous) accretion rates of the
  SMBH. We posit that sources with currently low values of the
  accretion rate relative to Eddington and whose host galaxies contain
  substantial amounts of molecular gas may lack an efficient mechanism
  to transport gas from $\sim$100 pc scales down to the SMBH.

In the future, further insight can come from an expanded
LLQSO sample containing a larger number of objects and spanning a wider range in
accretion rate relative to Eddington. This way, we can further investigate links between SMBH
accretion and molecular gas as well as X-ray photon index to further
investigate gas transport and accretion modes of these sources.

\end{itemize}

\begin{figure*}
  \centering 
	\hbox{
\includegraphics[width=9.0cm,angle=0]{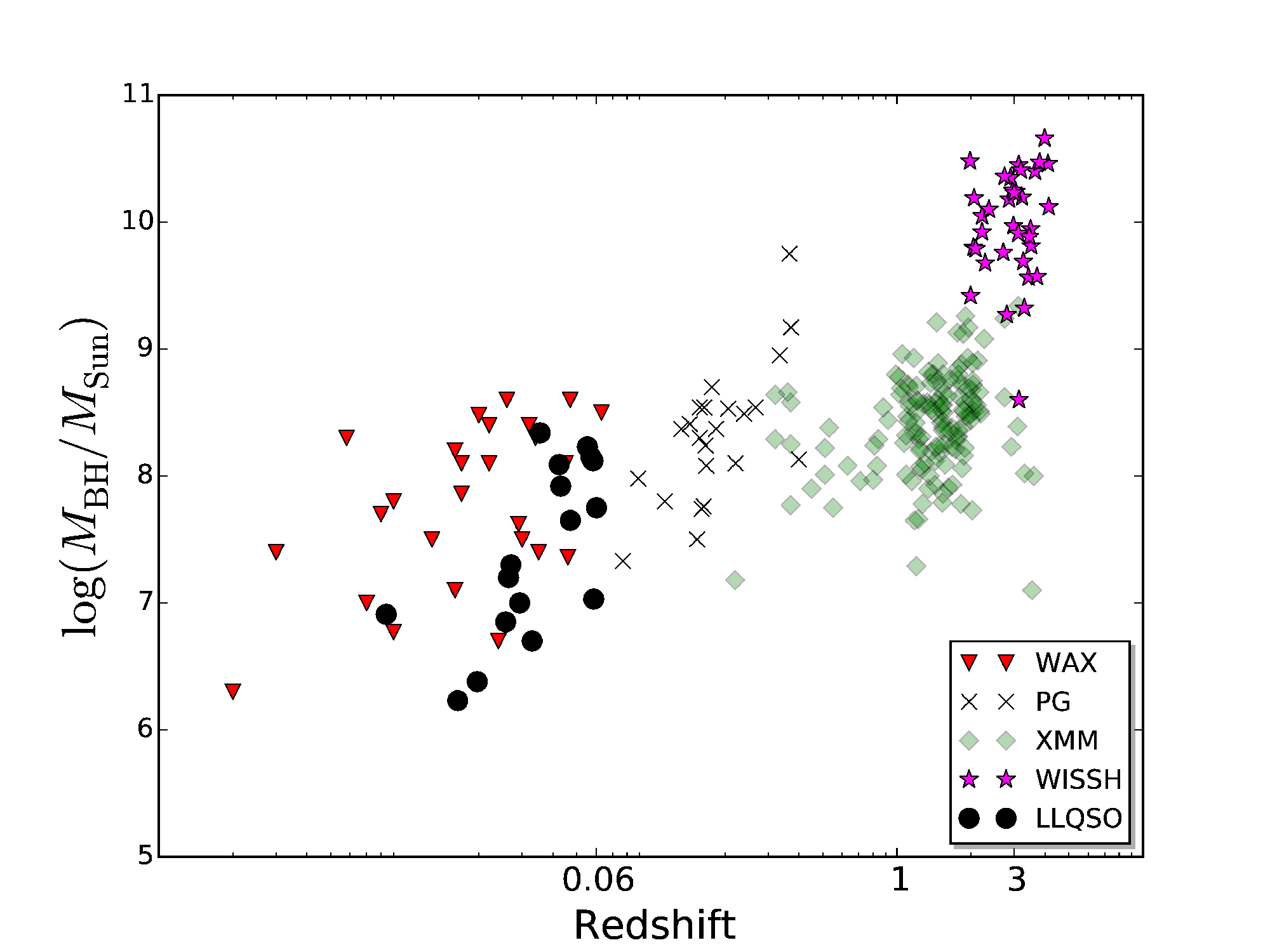} 
\includegraphics[width=9.0cm,angle=0]{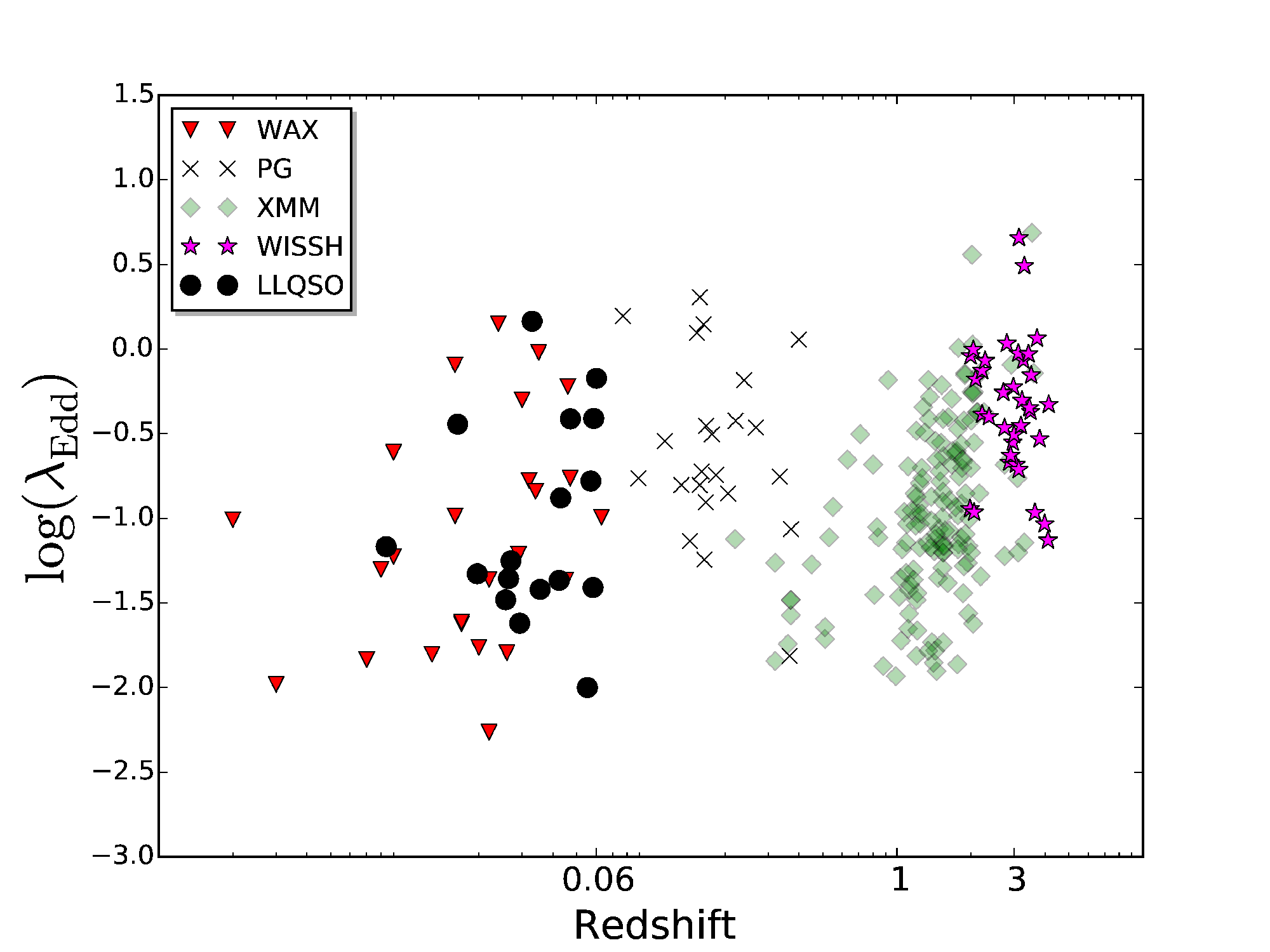} 
}
	\caption{ {\it Left:} The black hole mass of the LLQSO and the other samples plotted against redshift. Also see Tables \ref{Table:comparison} and \ref{Table:KS} and Section \ref{subsec:discussion1}. {\it Right:} Same as left, except for the Y axis, which is $\log\lambdaedd$. } \label{fig:evolution3}
\end{figure*}




\begin{table*}

{\footnotesize
\centering
  \caption{The list of sources in the LLQSO subsample. \label{Table:sources}}
  \begin{tabular}{lllllllllllllllll} \hline\hline \\ 

Id	& Source 	&Alternative name			&RA		&DEC		&Redshift	&Seyfert type		& {V-band}	&MBH			&$\nh^{\rm Gal}$&  \\
	&		&					&(J2000)	&(J2000)	&		&(optical class)	&magnitude	&$\log(\mbh/\msol)$	&($10^{20}\cmsqi$)&\\ \hline \\

1. 	&{HE0103-5842}			&ESO~113-G010		&16.325		&-58.437	&0.0257		&Sy1.8  	&14.59  	&6.85			&2.08& 	\\

2.	&{HE0203-0031}			&Mrk~1018		&31.566		&-0.291		&0.0424 	&Sy1.5  		&14.2  		&8.09			&2.43&	\\

3.	&HE0212-0059			&Mrk~590		&33.640		&-0.766		&0.0264		&Sy1.2 			&13.2  		&7.20			&2.65	&	\\

4.	&{HE0227-0913}			&Mrk~1044		&37.522		&-8.998		&0.0164		& NLSy-1 		& 14.5 		&6.23			&3.26	&	\\

5.	&HE0232-0900			&{NGC~0985}	&38.657		&-8.787		&0.043		& Sy 1.0	&13.8 		&7.92			&3.17&	\\

6.	&{HE0349-4036}			&Fairall~1116		&57.923		&-40.466	&0.0582		&Sy 1.0  	&14.99		&8.12			&2.45&\\

7.	&{HE0403-3719}			&ESO~359-G19		&61.256		&-37.187	&0.0552		&Sy 1.0  		&16.05		&8.23			&0.63 & \\

8.	&{HE0433-1028}			&Mrk~618		&69.092		&-10.375	&0.0355		&Sy 1.0	&15.0		&8.34			&4.61 &  \\

9.	&{HE0949-0122}			&Mrk~1239		&148.078	&-1.612		&0.0197		&Sy 1.5  		&13.3 		&6.38			&3.69 &  \\ 

10.	&{HE1011-0403}			&PG~1011-040		&153.586	&-4.311		&0.0586		&Sy 1.2 		&15.49		&7.03			&3.71 &  \\

11.	&HE1126-0407			&{Mrk~1298}		&172.319	&-4.402		&0.0601		&Sy 1.0 		&14.4		&7.75			&4.35 &  \\

12.	&HE1136-2304			&-			&174.713	&-23.360	&0.027		&Sy 1.0			&17.37		&7.30			&3.30 & \\

13.	&HE1143-1810			&-			&176.419	&-18.454	&0.0329		&Sy 1.5 		&15.0		&6.70			& 3.05 & \\

14.	&HE1237-0504$^2$		&NGC~4593		&189.914	&-5.344		&0.0084		&Sy 1.0 	&11.67		&6.91			& 1.89 &  \\

	  15. 	&{HE2129-3356}	   	&{CTS~A08.12}		&323.009	&-33.715	&0.0293		&Sy 1.2 		&15.7		&7.69			& 3.52 &  \\


16.	&HE2302-0857			&Mrk~0926		&346.181	&-8.685		&0.0471		&Sy1.5 			&14.6		&7.65			& 2.91  \\ 




\hline \hline

\end{tabular}  
{ The references for the black hole mass for the sources are listed in Appendix A.}
}
\end{table*}

\clearpage
\begin{table*}

{\footnotesize
\centering
  \caption{The X-ray observations of the LLQSO subsample. \label{Table:obs}}
  \begin{tabular}{llllllllllllllll} \hline\hline \\ 

Id		&Source		&X-ray		& obs-id	&Date of obs	& exposure	& \\ 
		&		&Satellite	&		&		&(ks)		& 	&\\ \hline \\

1. 		&ESO~113-G010	&\xmm{}		&0301890101	&10-11-2005	&104		\\
		&		&\xmm{}		&0103861601	&03-05-2001	&8		\\

2.		&Mrk~1018	&\suzaku	&704044010	&03-07-2009	&44				\\ 
		&		&\xmm		&0201090201	&15-01-2005	&12				\\ 
		&		&\xmm		&0554920301	&07-08-2008	&18				\\ 
		&		&\chandra{}	&18789		&25-02-2016	&30				\\ 
		&		&\chandra{}	&12868		&27-11-2010	&25				\\

3.		&Mrk~590	&\xmm{}		&0201020201	&04-07-2004	&113		\\
		&		&\xmm{}		&0109130301	&01-01-2002	&11		\\
		&		&\suzaku{}	&705043010	&23-01-2011	&62		\\
		&		&\suzaku{}	&705043020	&26-01-2011	&41		\\

4.		&Mrk~1044	&\xmm{}		&0695290101	&27-01-2013	&134		\\
		&		&\xmm{}		&0695290201	&27-01-2013	&42		\\
		&		&\xmm{}		&0112600301	&23-07-2002	&8		\\
		&		&\chandra{}	&18685		&20-09-2009	&14		\\

5.		&NGC~0985	&\xmm{}		&{0743830501}	&13-01-2015	&139		\\
		&		&\xmm{}		&0743830601	&25-01-2015	&122		\\
		&		&\xmm{}		&0150470601	&15-07-2003	&58		\\
		&		&\xmm{}		&0690870501	&10-08-2013	&104		\\
		&		&\xmm{}		&0690870101	&20-07-2013	&21		\\
		&		&\suzaku{}	&704042010	&15-07-2009	&32		\\
		&		&\chandra{}	&12866		&06-10-2010	&25		\\

6.		&Fairall~1116	&\xmm{}		&0301450301	&28-08-2005	&21		\\

7.		&ESO~359-G19	&\xmm{}		&0201130101	&09-03-2004	&24	\\

8.		&Mrk~618	&\xmm{}		&030700131	&15-02-2006	&18	\\

9.		&Mrk~1239	&\xmm{}		&0065790101	&12-11-2001	&10	\\ 
		&		&\suzaku{}	&06-05-2007	&06-05-2007	&63	\\ 

10.		&PG~1011-040	&\xmm{}		&0202060101	&08-05-2005	&32	\\

11.		&Mrk~1298	&\xmm{}		&0606150101	&21-06-2009	&134	\\
		&		&\xmm{}		&0202060201	&31-12-2004	&34	\\
		&		&\xmm{}		&0556230701	&15-06-2008	&31	\\
		&		&\xmm{}		&0556231201	&13-12-2008	&12	\\
		&		&\xmm{}		&0728180301	&12-06-2014	&23	\\
		&		&\xmm{}		&0728180401	&28-06-2014	&28	\\
		&		&\xmm{}		&0728180501	&14-06-2015	&18	\\

12.		&HE1136-2304	&\xmm{}		&0741260101	&02-07-2014	&110	\\

13.		&HE1143-1810	&\xmm{}		&0201130201	&08-06-2004	&34	\\
		&		&\chandra{}	&12873		&15-12-2010	&16	\\

14.		&NGC~4593	&\xmm{}		&0109970101	&02-07-2000	&28	\\ 
		&		&\xmm{}		&0740920501	&04-01-2015	&23	\\
		&		&\xmm{}		&0740920201	&29-12-2014	&26	\\
		&		&\xmm{}		&0740920601	&01-01-2015	&32	\\
		&		&\xmm{}		&0740920401	&02-01-2015	&26	\\
		&		&\xmm{}		&0740920301	&31-12-2014	&26	\\
		&		&\suzaku{}	&{702040010}	&15-12-2007	&119	\\

15. 		&CTS~A08.12	&\xmm{}		&0201130301	&30-10-2004	&46		\\


16.		&Mrk~0926	&\xmm{}		&0109130701	&01-12-2000	&12		\\ 
		&		&\xmm{}		&0109130901	&01-12-2000	&11		\\
		&		&\suzaku{}	&704032010	&02-12-2009	&11		\\

\hline \hline

\end{tabular}  


}
\end{table*}


\begin{table*}

{\footnotesize
\centering
  \caption{The UV and X-ray monochromatic fluxes and the $\alpha_{\rm OX}$ values of the sources. \label{Table:alphaox}}
  \begin{tabular}{lllllllllllllll} \hline\hline

Id	& Source 		&obsid		&$\log\rm F_{\rm 2500 \AA}^a$		&OM UV filter$\rm ^b$	& $\log\rm F_{\rm 2 \kev}$  		&$\rm A_{\lambda}^c$	& $\alpha_{\rm OX}$\\  	
	&			&		&$\funit {\rm \AA}^{-1}$	&			&$\funit {\kev}^{-1}$		&(Gal)			&			\\ \hline \\           	

1.	&ESO~113-G010		&0301890101	&$\rm -14.511$		&UVM2		&$-11.977$			&$0.212$		&$-1.233$		\\

2.	&Mrk~1018		&0554920301	&$\rm -14.892$		&UVM2		&$-11.481$			&$0.211$		&$-1.281$		\\

3. 	&Mrk~590		&0201020201	&$\rm -14.795$		&UVM2		&$-11.805$			&$0.285$		&$-1.063$	\\

4.	&Mrk~1044		&0112600301	&$\rm -14.163$		&V		&$-11.622$			&$0.269$		&$-1.235$	\\

5.	&NGC~0985		&0743830501	&$\rm -13.582$		&UVM2		&$-11.450$			&$0.256$		&$-1.391$	\\

6.	&Fairall1116		&0301450301	&$\rm -14.122$		&UVM2		&$-11.809$			&$0.095$		&$-1.311$	\\

7.	&ESO~359-G19		&0201130101	&$\rm -14.806$		&UVM2		&$-12.148$			&$0.042$		&$-1.175$	\\

8.	&Mrk~618		&030700131	&$\rm -13.903$		&UVM2		&$-11.475$			&$0.527$		&$-1.296$	\\

9.	&Mrk~1239		&0065790101	&$\rm -15.621$		&U		&$-13.261$			&$0.284$		&$-1.305$	\\

10.	&PG1011-040		&0202060101	&$\rm -13.835$		&UVM2		&$-13.408$			&$0.256$		&$-2.048$	\\

11.	&Mrk~1298		&0606150101	&$\rm -14.826$		&UVM2		&$-12.867$			&$0.400$		&$-1.853$	\\

12.	&HE1136-2304		&0741260101	&$\rm -14.688$		&UVM2		&$-11.605$			&$0.256$		&$-1.026$	\\

13. 	&HE1143-1810		&0201130201	&$\rm -13.606$		&UVM2		&$-11.078$			&$0.276$		&$-1.241$	\\

14.	&NGC~4593		&0109970101	&$\rm -14.000$		&UVW2		&$-10.959$			&$0.192$		&$-1.037$	\\

15.	&CTS A08.12		&0201130301	&$\rm -14.519$		&UVM2		&$-11.805$			&$0.383$		&$-1.176$	\\


16. 	&Mrk~0926		&0109130701	&$\rm -14.149$		&UVM2		&$-11.110$			&$0.288$		&$-1.044$	\\ \hline

\end{tabular}

{$^a$The UV flux of the LLQSOs measured in the observed frame.}\\
{$^b$ $U- 3440 \rm \AA$, $UVM2- 2310 \rm \AA$, $V- 5430 \rm \AA$. } \\ 
{$^c$ The Galactic extinction coefficient obtained from NED, at a wavelength corresponding to that of the OM UV filter used.}
}
\end{table*}


\begin{table*}

{\footnotesize
\centering
  \caption{The X-ray continuum spectral properties of LLQSO. \label{Table:xray1}}
  \begin{tabular}{llllllllllllllll} \hline\hline \\

	  Id	& 	Source 		&Obsid		&ztbabs		&power law	&bbody-1	&bbody-2	&{\it MYTorus}$^A$	&$\chi^2/dof\sim \chi^2_{\nu}$	&\\  
	  &			&		&$\nh$		&$\Gamma$	&$kT$		&$kT$		&$\nh$(inclination)			&		\\          	
	  &			&		&($\cmsqi$)	&		&($\kev$)	&($\kev$)	&			\\ \hline \\

1. 		&ESO~113-G010	&0301890101	&--		&$1.93\pm 0.05$	&$0.095\pm 0.002$&$0.282\pm 0.020$&--		&${230}/{241}\sim 0.95$			\\
		&		&0103861601	&--		&$1.79\pm 0.10$	&$0.101\pm 0.02$&$0.256\pm 0.07$  &		&${99}/{103}\sim 0.96$		\\ 	

 2.		&Mrk~1018	&0554920301	&--		&$2.04\pm 0.02$	&$0.092\pm 0.008$&--	&$51\pm 21$($<63$)	&$269/230\sim1.17$		\\
		&		&18789		&--		&$1.70\pm 0.05$	&$<0.018$	&--		&--		&$283/317\sim 0.88$		\\	
		&		&12868		&--		&$1.62\pm 0.11$	&$0.1116\pm 0.201$&--		&--		&$399/407\sim 0.98$		\\	
		&		&{704044010}	&--		&$1.87\pm 0.02$	&$0.097\pm 0.001$&-		&$93_{-29}^{+50} (<33)$	&$705/699\sim 1.01$		\\

3.		&Mrk~590	&0201020201	&--		&$1.67\pm 0.02$	&$0.136\pm 0.006$&--		&--		&$318/249\sim 1.28$	\\
		&		&0109130301	&--		&$1.71\pm 0.04$	&$0.133\pm 0.01$&--		&--		&$170/179\sim0.95$		\\ 	
		&		&{705043010}	&--		&1.70		&--		&--		&--		&$912/834 \sim 1.09$		\\
		&		&{705043020}	&--		&1.68		&--		&--		&--		&$568/551 \sim 1.03$		\\

4.		&Mrk~1044	&0695290101	&--		&$2.24\pm 0.02$	&$0.068\pm 0.005$&$0.147\pm 0.001$&--		&{$355/252\sim 1.40$}	\\ 
		&		&0112600301	&--		&$2.21\pm 0.04$	&$0.046\pm 0.003$&$0.107\pm 0.002$&--		&$179/174\sim1.02$		\\
		&		&18685		&--		&$1.86\pm0.07$		&$0.098\pm 0.002$&-		&--	&{ $392/320\sim 1.22$} \\

5.		&NGC~0985	&0743830501	&--		&$1.72\pm 0.03$	&$0.093\pm 0.001$	&$0.320\pm 0.020$&--		&{$375/253\sim 1.48$}	\\
		&		&0743830601	&--	 	&$1.81\pm 0.02$	&$0.091\pm 0.001$	&$0.314\pm 0.014$		&--		&{$473/254 \sim 1.86$}		\\ 	
		&		&0150470601	&--		&$1.55\pm 0.04$	&$0.098\pm0.003$	&$0.318\pm 0.051$&--		&$317/250 \sim 1.27$		\\	
		&		&0690870501	&--		&$1.24\pm 0.10$	&$0.093\pm 0.06$	&$0.259\pm 0.062$		&--	&{$560/248\sim 2.25$}		\\
		&		&0690870101	&--		&$1.34\pm 0.02$	&$0.085\pm 0.001$	&$0.105	\pm 0.042$	&--	&$293/224 \sim 1.31$	\\
		&		&12866		&--		&$1.42\pm 0.04$	&$0.120 \pm0.003$  	&$0.432\pm 0.052$	&--	&$372/362 \sim 1.03$	\\
		&		&{704042010}	&1.19e+21	&$1.77\pm 0.10$	&$0.058\pm 0.06$		&--		&--	&$761/714 \sim 1.07$		\\

6.		&Fairall~1116	&0301450301	&--		&$1.86\pm 0.05$		&$0.077\pm 0.019$&$0.187\pm 0.028$	&--	&$247/211\sim 1.17$	\\

7.		&ESO~359-G19	&0201130101	&--		&$1.70\pm 0.05$		&$0.126\pm 0.013$		&--	&--	&$214/192 \sim 1.12$	\\

8.		&Mrk~618	&030700131	&$<$1e+20	&$2.18\pm 0.04$		&$0.109\pm 0.091$		&--		&--	&$223/189 \sim 1.18$	\\

9.		&Mrk~1239	&0065790101	&--		&$<1.89$	&$0.132\pm 0.021$		&--		&--	&$22/20 \sim 1.11$	\\	
		&		&{702031010}&3.77e+21	&2.38		&0.094		&--		&$35_{-3.1}^{+4.4}(<42)$			&$377/264 \sim 1.43$ 	\\

10.		&PG~1011$-$040	&0202060101	&--		&$2.38\pm 0.18$	&$0.077\pm 0.011$&--		&--		   	&$86/77 \sim 1.12$&	\\

11.		&Mrk~1298	&0606150101	&--		&$2.13\pm 0.14$	&$0.088\pm 0.011$&--		&$11.8\pm 0.80$($<45$)		&$324/213 \sim 1.52$&	\\
		&		&0202060201	&--		&$2.17\pm0.05 $	&$0.084\pm 0.14$&--		&$3.4\pm 0.90$($<67$)		&$195/149 \sim 1.31$	\\
		&		&0556230701	&--		&$2.14\pm0.40$	&$0.097\pm 0.022$&--		&--			&$51/46 \sim 1.13$	\\
		&		&0556231201	&--		&$2.46\pm 0.22$	&$0.101\pm 0.031$&--		&--			&$89/78 \sim 1.15$	\\
		&		&0728180301	&--		&$1.97\pm 0.25$	&$0.096\pm 0.011$&--		&$3.4\pm 0.90$($<45$)		&$139/140 \sim 0.99$	\\
		&		&0728180401	&--		&$2.12\pm0.41$	&$0.096\pm 0.012$&--		&$5.4\pm 2.41$($<60$)		&$160/133 \sim 1.21$	\\
		&		&0728180501	&--		&$2.21\pm 0.18$	&$0.097\pm0.032$&--		&$2.7\pm 0.91$($<62$)		&$155/125 \sim 1.24$	\\

12.		&HE~1136$-$2304	&0741260101	&$(1.3\pm 0.1)$e+21&$1.52\pm 0.05$&$0.358\pm 0.028$&$0.138\pm 0.013$&--		&$310/257\sim 1.20$&\\

13.		&HE~1143$-$1810	&0201130201	&--		&$2.15\pm0.02$	&$0.097\pm 0.001$&-		&$23.3\pm3.3$($<90$)	 &$359/257 \sim 1.39$&\\
		&		&12873		&--		&$1.45\pm 0.03$		&$0.114\pm 0.012$	&-		&--		&{$442/415 \sim 1.07$}\\

14.		&NGC~4593	&0109970101	&--		&$1.75\pm 0.04$	&$0.093\pm 0.002$&$0.265\pm 0.022$&--		&$294/247\sim 1.19$	\\ 
		&		&0740920501	&--		&$1.61\pm 0.03$	&$0.101\pm 0.071$&$0.348\pm 0.013$&--		&$318/245 \sim 1.30$	\\
		&		&0740920201	&--		&$1.60\pm 0.07$	&$0.099\pm 0.011$&$0.335\pm 0.011$&--		&$298/247 \sim 1.21$	\\
		&		&0740920601	&--		&$1.63\pm 0.05$	&$0.102\pm 0.066$&$0.358 \pm 0.009$&--		&$269/250 \sim 1.08$	\\
		&		&0740920401	&--		&$1.47\pm 0.02$	&$0.103\pm 0.012$&$0.366\pm 0.011$&--		&$320/238 \sim 1.34$	\\
		&		&0740920301	&--		&$1.39\pm 0.04$	&$0.103\pm 0.012$&$0.372\pm 0.013$&--		&$277/244 \sim 1.13$	\\
		&		&{702040010}&--			&$1.53\pm 0.05$	&$0.054\pm 0.011$&--		&--		&$1805/1611 \sim 1.12$		\\

15. 		&CTS~A08	&0201130301	&--		&$1.39\pm 0.04$	&$0.109\pm 0.071$&$0.296\pm 0.030$&--		&$248/246\sim 1.00$& \\


16.		&Mrk~0926	&0109130701	&--		&$1.73\pm0.03$		&$0.122\pm 0.011$	&-		&--		&$247/243 \sim 1.02$& \\
		&		&{704032010}&-- 		&$1.84\pm 0.04$		&$0.083\pm 0.012$	&--		&$46_{-4.5}^{+6.2} (<20)$	&$2053/1593 \sim 1.29$ \\

\hline \hline

\end{tabular}  

}

{$^A$ The {\it MYTorus} $\nh$ column density in units of $10^{22}\cmsqi$. The inclination angle in the brackets are expressed in degrees. }\\



\end{table*}


\begin{table*}

{\footnotesize
\centering
  \caption{The discrete X-ray spectral properties of LLQSO. \label{Table:xray2}}
  \begin{tabular}{llllllllllllllll} \hline\hline \\

Id		& Source 	&obsid		&FeK		&FeK$^b$		&Diskline	&WA1		&WA1		&WA2		&WA2		\\  
		&		&		&line E(EQW)$^a$	&Line E	(EQW)$^a$	&Line E (EQW)$^a$	&$\log\xi$	&$\log\nh$	&$\log\xi$	&$\log\nh$			\\          	
		&		&		&$\kev$ 	&$\kev$		&$\kev$(eV)	&($\xiunit$)	&($\cmsqi$)	&($\xiunit$)	&($\cmsqi$)	\\ \hline \\

	  1. 		&ESO~113-G010	&0301890101	&6.41(65)&7.00(107)	&--		&$3.31\pm 0.16$	&$21.82\pm 0.17$&--		&--		\\	
		&		        &0103861601	&6.41(43)&7.15(49)	&--		&$2.51\pm 0.22$	&$21.74\pm 0.42$&--		&--		\\

2.		&Mrk~1018	&0554920301	&--		&--		&--		&--		&--		&--		&--		\\	
		&		&18789		&--		&--		&--		&--		&--		&--		&--		\\	
		&		&12868		&--		&--		&--		&--		&--		&--		&--		\\	
		&		&{704044010}	&6.37		&--		&--		&--		&--		&--		&--		\\

	  3.	&Mrk~590	&0201020201	&6.41(131)	&--		&--		&--		&--		&--		&--			\\	
		&		&0109130301	&6.41(245)	&--		&--		&--		&--		&--		&--			\\
		&		&{705043010}	&6.42		&--		&--		&--		&--		&--		&--	\\
		&		&{705043020}	&6.43		&--		&--		&--		&--		&--		&--	\\

	  4.	&Mrk~1044$^1$	&0695290101	&6.81(103)	&--		&6.96$^1$(680)	&--		&--		&--		&--			\\	
		&		&0112600301	&6.47(162)	&--		&6.92(1322)	&--		&--		&--		&--			\\
		&		&18685		&6.68(219)	&--		&--		&--		&--		&--		&--			\\

	  5.	&NGC~0985	&0743830501	&6.44(87)	&--		&--		&$2.07\pm 0.06$	&$21.62\pm 0.05$&$3.16\pm 0.05$	&$22.11\pm 0.06$			\\	
		&		&0743830601	&6.44(82)	&--		&--		&$2.35\pm 0.05$	&$21.56\pm 0.07$&$3.14\pm 0.04$	&$22.12\pm 0.07$			\\
		&		&0150470601	&6.42(162)	&--		&--		&$2.08\pm 0.11$	&$21.75\pm 0.10$&$3.02\pm 0.10$	&$22.10\pm 0.06$			\\
		&		&0690870501	&6.46(152)	&--		&--		&$2.06\pm 0.11$	&$22.09\pm 0.12$&$3.38\pm 0.10$	&$22.59\pm 0.08$			\\
		&		&0690870501	&6.42(153)	&--		&--		&$2.19\pm 0.11$	&$22.2\pm 0.12$	&$2.96\pm 0.10$	&$22.65	\pm 0.06$			\\
		&		&12866		&6.33(105)	&--		&--		&$2.42\pm 0.12$	&$21.78\pm 0.12$&$3.11\pm 0.10$	&$22.23	\pm 0.08$		\\
		&		&{704042010}	&6.39		&--		&--		&--		&--		&--		&--		\\

6.		&Fairall~1116	&0301450301	&6.47(169)	&--		&--		&--		&--		&--		&--				\\

	  7.		&ESO~359-G19	&0201130101	&6.40(257)		&7.08(128)	&--		&--		&--		&--		&--			\\

8.		&Mrk~618	&030700131	&--		&--		&6.58(474)	&--		&--		&--		&--			\\

9.		&Mrk~1239	&0065790101$^2$	&5.14		&--		&--		&$0.75\pm0.30$	&$21.67\pm 0.30$	&--		&--			\\
		&		&{ 702031010}	&6.64		&--		&--		&1.09		&21.77		&--		&--		\\

10.		&PG~1011-040	&0202060101	&--		&--		&--		&--		&--		&--		&--				\\

11.		&Mrk~1298	&0606150101	&6.28(249)	&--		&--		&$2.15\pm 0.12$		&$22.35\pm 0.11$&--		&--			\\
		&		&0202060201	&6.40(150)	&--		&--		&$0.93\pm0.31 $		&$21.99\pm 0.22$&--		&--			\\
		&		&0556230701$^2$	&4.86(145)	&--		&--		&$2.78\pm0.99$		&$22.74\pm 0.51$&--		&--			\\
		&		&0556231201	&6.11(111)	&--		&--		&$1.99\pm 0.41$		&$22.3\pm 0.22$	&--		&--			\\
		&		&0728180301	&6.32(340)	&--		&--		&$2.10\pm 0.61$		&$21.90\pm 0.77$&--		&--			\\
		&		&0728180401	&6.32(570)	&--		&--		&$2.25\pm0.22$		&$22.34\pm0.42$	&--		&--			\\
		&		&0728180501	&6.29(374)	&--		&--		&$2.29\pm0.32$		&$22.24\pm0.55$	&--		&--			\\

12.		&HE1136-2304	&0741260101	&6.45(65)	&6.93(28)	&--		&--		&--		&--		&--			\\

13.		&HE~1143-1810	&0201130201	&6.41(19)	&--		&--		&--		&--		&--		&--			\\
		&		&12873		&6.41(17)	&--		&--		&--		&--		&--		&--			\\

14.		&NGC~4593	&0109970101	&6.40(102)	&--		&--		&$2.18\pm 0.21$	&$21.09\pm0.21$		&$3.08\pm 0.17$ &$21.58\pm 0.14$			\\
		&		&0740920501	&6.40(99)	&--		&--		&$1.24\pm0.15 $	&$20.65\pm0.11$		&$2.89\pm0.13$	&$21.27\pm 0.19$			\\
		&		&0740920201	&6.40(81)	&--		&--		&$2.12\pm0.30$	&$20.98\pm 0.15$	&$3.07\pm 0.19$	&$21.52\pm 0.15$			\\
		&		&0740920601	&6.40(94)	&--		&--		&$1.10\pm 0.19$	&$20.64\pm 0.27$	&$3.04\pm 0.13$	&$21.26\pm 0.08$			\\
		&		&0740920401	&6.40(169)	&--		&--		&$2.28\pm 0.12$	&$21.05\pm 0.23$	&$3.17\pm 0.22$	&$21.18\pm 0.17$			\\
		&		&0740920301	&6.40(133)	&--		&--		&$2.77\pm0.13$	&$20.62\pm 0.11$	&$2.93\pm 0.21$	&$21.95\pm 0.15$			\\
		&		&{ 702040010}	&6.40		&7.00		&--		&--  	   	&--			&--		&--		\\

15. 		&CTS~A08	&0201130301	&6.4(94)	&--		&--		&--		&--		&--		&--			\\


16.		&Mrk~0926	&0109130701	&6.38(90)	&6.70(43)	&--		&--		&--		&--		&--			\\
		&		&{704032010}	&6.41		&--		&--		&--		&--		&--		&--			\\




\hline \hline

\end{tabular}  

}

{$^1$ The line energy of the diskline profile is pegged at $6.96 \kev$.  }\\

{$^2$ The data quality being poor, the Fe K line could not be constrained.}\\

{$^a$ The bracketed quantities are the Fe K equivalent widths in $\ev$.}\\

{$^b$ This column lists the parameters for the higher ionized Fe emission line}

\end{table*}


\begin{table*}
{\footnotesize
\centering
  \caption{The flux and luminosity of LLQSO. \label{Table:xray3}}
  \begin{tabular}{llllllllllllllll} \hline\hline \\

	  Id		& Source 	&obsid		&$\rm \log F^1_{0.3-2 \kev}$	&$\rm \log F^1_{2-10 \kev}$	&$\rm \log L^1_{0.3-2 \kev}$	&$\rm \log  L^1_{2-10 \kev}$	&$\log \lbol$	&$\log \ledd ^1$	&$\kappa_{\rm Lbol}$	&$\lambda_{\rm Edd}^2$		\\  
		&		&		&$\funit$	&$\funit$	&$\lunit$	&$\lunit$	&$\lunit$	&$\lunit$& 		\\ \hline \\

	  1. 	&ESO~113-G010	&0301890101	&$-11.216\pm0.007$	&$-11.484\pm0.002$	&$42.88\pm0.02$		&$42.62\pm 0.02$	&$43.76$	&44.96	&13.80	&0.063	\\	
		&		&0103861601	&$-11.359\pm0.011$	&$-11.588\pm0.012$	&$42.74\pm0.03$		&$42.51\pm0.04$		&$43.65$	&44.96	&13.80	&0.049	\\

	  2.	&Mrk~1018	&0554920301	&$-10.848\pm0.008$	&$-10.927\pm0.008$	&$43.69\pm0.03$		&$43.61\pm0.05$		&$44.81$	&46.20	&15.81		&0.041	\\	
		&		&18789		&$-11.070\pm0.007$	&$-11.961\pm0.007$	&$43.47\pm0.02$		&$42.58\pm0.06$		&$43.78$	&46.20	&15.81		&0.003	\\	
		&		&12868		&$-11.713\pm0.006$	&$-11.512\pm0.006$	&$42.18\pm0.02$		&$43.03\pm0.02$		&$44.23$	&46.20	&15.81		&0.010	\\
		&		&{ 704044010}&$-10.920\pm0.007$	&$-10.980\pm0.005$	&$43.62\pm0.05$		&$43.56\pm0.02$		&$44.76$	&46.20	&15.81		&0.036	\\

	  3.	&Mrk~590	&0201020201	&$-11.302\pm 0.009$	&$-11.176\pm 0.004$	&$42.81\pm0.04$		&$42.94\pm0.02$		&$43.91$	&45.31	&9.47		&0.039	\\	
		&		&0109130301	&$-11.411\pm 0.007$	&$-11.314\pm0.005$	&$42.70\pm0.02$		&$42.79\pm0.04$		&$43.77$	&45.31	&9.47		&0.028	\\	
		&		&{705043010}	&$-11.290\pm 0.009$	&$-11.130\pm 0.008$	&$42.82\pm0.08$		&$42.98\pm0.02$		&$43.95$	&45.31	&9.47		&0.044	\\
		&		&{705043020}	&$-11.340\pm 0.007$	&$-11.17\pm 0.006$	&$42.77\pm0.05$		&$42.94\pm0.06$		&$43.91$	&45.31	&9.47		&0.040	\\

	  4.	&Mrk~1044	&0695290101	&$-10.535\pm 0.007$	&$-10.965\pm 0.008$	&$43.14\pm0.02$		&$42.72\pm0.02$		&$43.85$	&44.34	&13.88		&0.327	\\	
		&		&0112600301	&$-10.709\pm 0.003$	&$-11.164\pm 0.006$	&$42.97\pm0.04$		&$42.51\pm0.06$		&$43.66$	&44.34	&13.88		&0.207	\\	
		&		&18685		&$-11.472\pm 0.008$	&$-11.622\pm 0.005$	&$42.20\pm0.06$		&$42.05\pm0.05$		&$43.20$	&44.34	&13.88		&0.072	\\

	  5.	&NGC~0985	&0743830501	&$-10.769\pm 0.008$	&$-10.885\pm 0.008$	&$43.79\pm0.02$		&$43.67\pm0.02$		&$45.03$	&46.03	&22.67		&0.100	\\	
		&		&0743830601	&$-10.591\pm 0.003$	&$-10.764\pm 0.008$	&$43.97\pm0.02$		&$43.79\pm0.02$		&$45.15$	&46.03	&22.67		&0.132	\\	
		&		&0150470601	&$-11.053\pm 0.003$	&$-10.984\pm 0.008$	&$43.51\pm0.02$		&$43.57\pm0.02$		&$44.93$	&46.03	&22.67		&0.079	\\	
		&		&0690870501	&$-11.367\pm 0.012$	&$-11.028\pm 0.008$	&$43.19\pm0.05$		&$43.53\pm0.07$		&$44.89$	&46.03	&22.67		&0.072		\\	
		&		&0690870501	&$-11.345\pm 0.010$	&$-10.991\pm 0.008$	&$43.21\pm0.04$		&$43.57\pm0.07$		&$44.92$	&46.03	&22.67		&0.078		\\
		&		&12866		&$-11.591\pm 0.006$	&$-11.551\pm 0.008$	&$42.97\pm0.04$		&$43.01\pm0.08$		&$44.36$	&46.03	&22.67		&0.021		\\	
		&		&{704042010}	&$-10.140\pm 0.007$	&$-10.850\pm 0.008$	&$44.42\pm0.07$		&$43.71	\pm0.04$	&$45.06$	&46.03	&22.67		&0.108		\\

	  6.	&Fairall~1116	&0301450301	&$-11.113\pm 0.002$	&$-11.265\pm 0.004$	&$43.74\pm0.02$		&$43.58\pm0.02$		&$44.82$	&46.23	&17.33		&0.039		\\

	  7.	&ESO~359-G19	&0201130101	&$-11.637\pm 0.008$	&$-11.519\pm 0.008$	&$43.16\pm0.02$		&$43.28\pm0.02$		&$44.35$	&46.34	&11.92		&0.010		\\

	8.	&Mrk~618	&030700131	&$-10.751\pm 0.008$	&$-11.027\pm 0.008$	&$44.09\pm0.05$		&$43.82\pm0.05$		&$45.04$	&46.45	&16.54		&0.038			\\

	9.	&Mrk~1239	&0065790101	&$-11.811\pm 0.003$	&$-12.369\pm 0.008$	&$42.13\pm0.06$		&$41.57\pm0.07$		&$42.80$	&44.49	&17.01		&0.020		\\
		&		&702031010	&$-11.590\pm 0.005$	&$-12.000\pm 0.005$	&$42.35\pm0.06$		&$41.94\pm0.08$		&$43.17$	&44.49	&17.01		&0.047		\\

	10.	&PG~1011-040	&0202060101	&$-12.537\pm 0.007$	&$-13.044\pm 0.008$	&$42.33\pm0.02$		&$41.82\pm0.02$		&$44.73$	&45.14	&809		&0.389		\\

	11.	&Mrk~1298	&0606150101	&$-12.264\pm 0.011$	&$-11.929\pm 0.008$	&$42.65\pm0.03$		&$42.99\pm0.02$		&$45.33$		&45.86	&217.26		&0.293			\\	
		&		&0202060201	&$-12.595\pm 0.009$	&$-11.923\pm 0.009$	&$42.32\pm0.03$		&$42.99\pm0.02$		&$45.33$		&45.86	&217.26		&0.297			\\	
		&		&0556230701	&$-12.229\pm 0.010$	&$-11.761\pm 0.010$	&$42.69\pm0.04$		&$43.16\pm0.03$		&$45.49$		&45.86	&217.26		&0.431		\\
		&		&0556231201	&$-11.746\pm 0.009$	&$-11.568\pm 0.009$	&$43.17\pm0.04$		&$43.35\pm0.03$		&$45.69$		&45.86	&217.26		&0.672			\\	
		&		&0728180301	&$-12.568\pm 0.008$	&$-11.862\pm 0.008$	&$42.35\pm0.05$		&$43.06\pm0.03$		&$45.39$		&45.86	&217.26		&0.342		\\	
		&		&0728180401	&$-12.411\pm 0.007$	&$-12.018\pm 0.009$	&$42.51\pm0.07$		&$42.90\pm0.04$		&$45.24$		&45.86	&217.26		&0.238		\\	
		&		&0728180501	&$-12.376\pm 0.011$	&$-11.832\pm 0.007$	&$42.54\pm0.03$		&$43.09\pm0.02$		&$45.42$		&45.86	&217.26		&0.366		\\

	12.	&HE1136-2304	&0741260101	&$-11.112\pm 0.012$	&$-10.985\pm 0.008$	&$43.08\pm0.05$		&$43.21\pm0.06$		&$44.16$		&45.41	&8.92		&0.056			\\

	13.	&HE~1143-1810	&0201130201	&$-10.364\pm 0.013$	&$-10.539\pm 0.007$	&$44.01\pm0.02$		&$43.83\pm0.02$		&$44.98$		&44.81	&14.11		&1.46		\\	
		&		&12873		&$-11.581\pm 0.009$	&$-11.261\pm 0.008$	&$42.79\pm0.02$		&$43.11\pm0.02$		&$44.25$		&44.81	&14.11		&0.279		\\

	14.	&NGC~4593	&0109970101	&$-10.331\pm 0.008$	&$-10.399\pm 0.008$	&$42.96\pm0.02$		&$42.90\pm0.02$		&$43.85$		&45.02	&9.07		&0.068			\\	
		&		&0740920501	&$-10.556\pm 0.007$	&$-10.552\pm 0.007$	&$42.73\pm0.04$		&$42.74	\pm0.04$	&$43.70$		&45.02	&9.07		&0.048		\\	
		&		&0740920201	&$-10.679\pm 0.009$	&$-10.629\pm 0.009$	&$42.62\pm0.05$		&$42.67\pm0.05$		&$43.62$		&45.02	&9.07		&0.040		\\	
		&		&0740920601	&$-10.579\pm 0.011$	&$-10.576\pm 0.009$	&$42.72\pm0.05$		&$42.72\pm0.05$		&$43.68$		&45.02	&9.07		&0.045		\\	
		&		&0740920401	&$-10.957\pm 0.012$	&$-10.823\pm 0.008$	&$42.34	\pm0.06$	&$42.47\pm0.06$		&$43.43$		&45.02	&9.07		&0.025		\\	
		&		&0740920301	&$-11.017\pm 0.010$	&$-10.839\pm 0.008$	&$42.28\pm0.07$		&$42.46\pm0.07$		&$43.41$		&45.02	&9.07		&0.025		\\	
		&		&{702040010}	&$-11.021\pm 0.008$	&$-10.990\pm 0.007$	&$42.27\pm0.08$		&$42.30\pm0.08$		&$43.26$		&45.02	&9.07		&0.017	\\

15. 		&CTS~A08	&0201130301	&$-11.328\pm 0.008$	&$-11.105\pm 0.008$	&$42.89\pm0.02$		&$43.12\pm0.02$		&$44.19$		&45.80	&11.95		&0.249			\\


16.		&Mrk~0926	&0109130701	&$-10.595\pm 0.010$	&$-10.507\pm 0.008$	&$44.03\pm0.07$		&$44.12\pm0.02$		&$45.08$		&45.76	&9.177		&0.209			\\	
		&		&{704032010}	&$-10.310\pm 0.011$	&$-10.240\pm 0.009$	&$44.32\pm0.08$		&$44.39\pm0.09$		&$45.35$		&45.76	&9.177		&0.387	\\




\hline \hline

\end{tabular}  

}

{$^1$The flux and luminosity are quoted in log units.}\\
{$^2$ $\lambda_{\rm Edd}=\lbol/\ledd$. }\\

\end{table*}


\begin{table*}

{\footnotesize
\centering
  \caption{The molecular hydrogen estimates of the LLQSO. \label{Table:mol}}
  \begin{tabular}{llllllllllllllllll} \hline\hline \\

Id	& Source 		&$L_{\rm CO}^{1}$	&M($\rm H_2$)	\\  
	&			&			&($10^9 \msol$)		&	\\ \hline \\

1. 	&ESO~113-G010		&-		&-				\\

2.	&Mrk~1018		&$<1.6$		&$<0.6$				\\

3.	&Mrk~590		&$4.7$		&$1.9$				\\

4.	&Mrk~1044		&$1.0$		&$0.4$					\\

5.	&NGC~0985		&$24.3$		&$9.7$				\\

6.	&Fairall~1116		&-		&					\\

7.	&ESO~359-G19		&-		&					\\

8.	&Mrk~618		&$22.6$		&$9.0$				\\

9.	&Mrk~1239		&$0.9$		&$0.4$					\\ 

10.	&PG~1011-040		&$8.9$		&$3.5$					\\

11.	&Mrk~1298		&$10.3$		&$4.1$				\\

12.	&HE1136-2304		&-		&-					\\

13.	&HE1143-1810		&-		&-					\\

14.	&NGC~4593		&$1.0$		&$0.4$					\\

15. 	&CTS~A08		&-		&-					\\


16.	&Mrk~0926		&$9.7$		&$3.9$					\\ 


\hline \hline

\end{tabular}  

{$^1$ CO(1-0) emission line luminosity obtained from \citet{2007A&A...470..571B}, in units of $10^8 \times \rm K\, km\, s^{-1}\, pc^2$.}

}
\end{table*}


\begin{table*}

{\footnotesize
\centering
  \caption{The average and standard deviation of the UV and X-ray properties of the sources. \label{Table:comparison}}
  \begin{tabular}{lllllllllllllll} \hline\hline

	  Samples	&$\log\lhard$	&$\log\lbol$	&$\log L_{\rm 2500 \AA}$	&$\alpha_{\rm OX}$	&$\log(\mbh/\msol)$		& $\log(\lambda_{\rm Edd})$ \\ 
	  		&$\lunit$	&$\lunit$	&$\lunit \, \rm Hz^{-1}$&			&				\\  \hline \\

	  LLQSO		&$43.10\pm0.68$	&$44.41\pm0.66$	&$29.60\pm0.93$		&$-1.29\pm0.27$		&$7.35\pm0.65$		&$-0.95\pm0.56$	\\
	  
	  WAX		&$43.30\pm0.74$	&$44.69\pm0.77$	&$29.39\pm0.97$		&$-1.08\pm0.29$		&$7.77\pm0.62$		&$-1.19\pm0.77$	\\

	  PG Quasars	&$44.22\pm0.53$	&$45.85\pm0.43$	&$30.21\pm0.42$		&$-1.53\pm0.14$		&$8.32\pm0.53$		&$-0.58\pm 0.50$		\\	

	  XMM-COSMOS	&$44.04\pm0.53$	&$45.48\pm0.63$	&$29.51\pm0.77$		&$-1.36\pm0.18$		&$8.41\pm0.39$		&$-0.96\pm 0.50$		\\

	  WISSH		&$45.44\pm0.41$	&$47.74\pm 0.19$&$32.25\pm0.17$		&$-1.80\pm 0.14$	&$9.98\pm0.43$		&$-0.35\pm0.39$		\\ \hline \

\end{tabular}

}
\end{table*}


\begin{table}
{\footnotesize
\centering
  \caption{The KS test results between the parameters of LLQSO and other quasar samples$\rm ^A$ \label{Table:KS}}
  \begin{tabular}{lllllllllllllll} \hline\hline

	  Quantities	&WAX		&PG-Quasars		&XMM-COSMOS		&WISSH	 			\\  \hline \\

	  LLQSO $\lhard$&(\textbf{0.17, 0.92}$\rm ^B$)	&(0.72, $4.06\times10^{-5}$)	&(0.66, $9.93\times10^{-7}$)	&(1.0, $7.26\times10^{-11}$)				\\
	  
	  LLQSO $\lbol$	&(\textbf{0.28, 0.34})	&(0.96, $1.05\times10^{-8}$)	&(0.71, $1.04\times10^{-7}$)	&(1.0, $7.26\times10^{-11}$) 			\\

	  LLQSO $L_{\rm 2500 \AA}$&(\textbf{0.26, 0.41})	&({0.5, $8\times 10^{-3}$})		&(\textbf{0.28, 0.13})		&(1.0, $7.26\times10^{-11}$)	\\	

	  LLQSO $\alpha_{\rm OX}$&(\textbf{0.39, 0.07})	&(0.83, $1.1\times 10^{-6}$)	&(0.46, $1.8\times 10^{-3}$)		&(1.0, $7.26\times10^{-11}$)		\\

	  LLQSO $\log(\mbh/\msol)$&(\textbf{0.37, 0.10}) &(0.56, $3\times 10^{-3}$)		&(0.65, $3.04\times10^{-6}$)	&(1.0, $7.26\times10^{-11}$)		\\
	  
	  LLQSO $\log(\lambda_{\rm Edd})$&(\textbf{0.28, 0.34})		&(\textbf{0.48, 0.02})		&(\textbf{0.22, 0.46})	&(0.56, $9.95\times10^{-4}$)		\\ \hline \

\end{tabular}
{$\rm ^A$ The bracketed quantities are output from KS test between a given parameter distribution of LLQSO and the corresponding sample listed at the top of the column.}\\
{$\rm ^B$ Quantities for which the null hypothesis in the KS test cannot be ruled out at a confidence $\ge 99\%$ are written in bold, implying that these samples have been likely derived from the same parent sample. See Section \ref{subsubsection:distributions} for a discussion.}
  
}
\end{table}


\appendix

\clearpage

\section{Previous studies of the sources in the LLQSO sample.}

\noindent 1. {\bf ESO~113$-$G010}:   This galaxy has been optically classified as a Seyfert 1.8
by \citet{1998A&A...333...48P}. However, later studies of the X-ray spectrum
by \citet{2012A&A...542A..30M} found no signature of neutral
absorption as is commonly detected for Seyfert 1.8 galaxies. Those authors
detected a large Balmer decrement ($\rm H\alpha/H\beta \sim 8$) in the optical continuum,
indicating a large amount of reddening in the optical. However, no corresponding
absorbing component has been detected in the UV or the X-rays. Those authors
conclude that the Balmer decrement detected in the optical could be
due to dusty warm absorbers where the dust does not affect the UV and
X-ray photons. The
black hole mass was obtained from \citet{2013ApJ...764L...9C} who have
used the relation between $L(5100\rm \AA)$ and $L_{\rm
H\alpha}$ \citep{2005ApJ...627..721G}.

In our work, we found $\alpha_{\rm OX}=1.23$, consistent with the
value calculated by \citet{2012A&A...542A..30M}. The baseline model
provides a good fit to the spectra. There are two gaussian components
necessary to model Fe K emission lines, one at a higher energy,
E=$7.00\pm 0.01$ keV, 
possibly indicating an Fe~XXVI emission line.  
\\

\noindent 2. {\bf Mrk~1018}: This source is classified as changing look AGN. \citet{1986ApJ...311..135C} 
discovered that the source had changed from Type 2 to Type 1 between 1979
and 1984. More recently, \citet{2016A&A...593L...8M} found that the
broad emission lines of the source had disappeared between 2009 and
2015. According to those authors, a decrease in the accretion rate was
possibly responsible for such a behaviour; however, they could not
definitively rule out variable
obscuration.  \citet{2013MNRAS.428.2901W} have studied this source
using \suzaku{} data and classified it as a `bare Seyfert galaxy' with
no X-ray absorption along the line of sight.

We did not detect any source photons in the latest 2015 \chandra{}
observation of this source, consistent with \citet{2016A&A...593L...8M}. We studied X-ray spectra taken between
2005 and 2010, when the source was in a Type 1
  state. We found that the source flux was mostly stable in X-rays during
  that time span, with a variation in $0.3-10\kev$ luminosity of $\le
  20\%$. The black hole mass for this source was obtained
  by \citet{2002ApJ...579..530W} using the observed stellar velocity
  dispersion.  \\

\noindent 3. {\bf Mrk~590}:  In our work the baseline model provides a good description of the spectra in all
the \xmm{} and \suzaku{} observations. The black hole mass for this
source has been obtained by reverberation mapping
by \citet{2000ApJ...533..631K}.
\\

\noindent 4. {\bf Mrk~1044}:  The black hole mass of this source has been estimated by \citet{2001A&A...377...52W} using the monochromatic continuum luminosity at $5100 \rm \AA$.  \\

\noindent 5. {\bf NGC~0985}:  This source is also known as Mrk~1048. \citet{2009ApJ...690..773K}  and \citet{2016A&A...586A..72E} have detected ionized absorption in X-rays in this source, similar to our findings. The black hole mass has been obtained using the $\rm H\beta$ emission line width \citep{2011ApJ...739...57K}. \\

\noindent 6. {\bf Fairall~1116}: \citet{2008A&A...482..499D} and \citet{2016A&A...588A..70B} 
have studied this source in samples of Seyfert galaxies and have
derived values of X-ray luminosity similar to ours. The black hole
mass of the source has been obtained using the $\rm H\beta$ emission
line width \citet{2003ApJ...583..124S}. \\

\noindent 7. {\bf ESO~359$-$G19}: \citet{2011A&A...530A.125C} and \citet{2016A&A...588A..70B} have 
studied this source in samples of Seyfert galaxies; we obtain similar X-ray spectral parameters
as those authors.The black hole mass of this source
has been obtained from \citet{2010ApJS..187...64G} using the $\rm
H\beta$ emission line width. \\

\noindent 8. {\bf Mrk~618}: The X-ray spectra has been studied by \citet{2011MNRAS.413.1206B} as a part of 126 sources selected on the basis of their 12 $\mu$m luminosity. The black hole mass for this source has been obtained from stellar velocity dispersion relation \citet{2002ApJ...579..530W}.  \\

\noindent 9. {\bf Mrk~1239}: The black hole mass of this source has been obtained from \citet{2001A&A...377...52W} using the monochromatic continuum luminosity at $5100 \rm \AA$. \\

\noindent 10. {\bf PG~1011$-$040}: The black hole mass of this source has been obtained from \citet{2001A&A...377...52W} using the monochromatic continuum luminosity at $5100 \rm \AA$. \\

\noindent 11. {\bf MRK~1298}: The black hole mass of this source has been obtained from \citet{2006ApJ...641..689V} using the $\rm H\beta$ emission line width.  \\

\noindent 12. {\bf HE~1136$-$2304}: The source is classified as a changing look Seyfert galaxy by 
\citet{2006ApJ...641..689V}, where the authors note that the source had changed its nature from 
Type 2 in 1993 to Type 1.5 in 2004, with the broad emission line
intensity greatly increased. On the other hand the X-ray spectral
observation with \xmm{} and \nustar{} have revealed only moderate
obscuration by intervening neutral gas, similar to our
findings. However, the authors could not definitively attribute the
changing look nature to its changing obscuration or change in
accretion rates. The black hole mass of this source has been obtained
from \citet{2006ApJ...641..689V} using the $\rm H\beta$ emission line
width. \\

\noindent 13. {\bf HE~1143-1810}:  This source has been studied as a part of 
several samples in
X-rays \citep{2007MNRAS.382..194N,2011A&A...530A.125C,2012ApJ...745..107W,2016A&A...588A..70B}
and it shows signs of a broad Fe line as well as the presence of warm
absorbers, similar to our findings.  The black hole mass is obtained
from \citet{2009ApJ...690.1322W}.\\

\noindent 14. {\bf NGC~4593}:  \citet{2012MNRAS.426.2522P,2014MNRAS.441.2613L} have studied this source in samples of Seyfert galaxies using \suzaku{} and \xmm{} and could not detect a narrow Fe $\rm K\alpha$ emission line along with warm absorption features, similar to our analysis. The black hole mass for this object is obtained from \citet{1999ApJ...516..672H} using reverberation mapping. \\

\noindent 15. {\bf CTS~A08}: The black hole mass of this source is obtained from \citet{2014A&A...561A.140B}. \\


\noindent 16. {\bf Mrk~0926}: This source is also known as MCG$-$2-58-22. The \suzaku{} observation of this source has been studied by \citet{2011ApJ...732...36R} where the authors could put a tight constraint on the reflection component, similar to our work. The black hole mass was obtained from \citet{2011AAS...21840804J} using the $\rm H\beta$ emission line width.
\\


\clearpage

\section{The X-ray spectral fits of the LLQSO}

\begin{figure*}
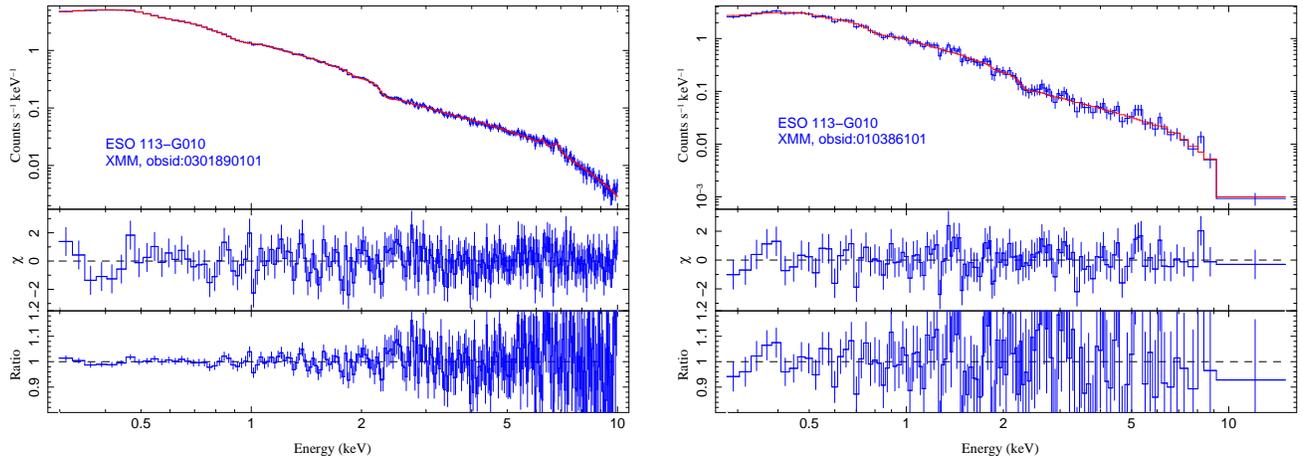

  \centering 

\hbox{
\includegraphics[width=6.0cm,angle=-90]{ESO113-G010_0301890101_multiplot.ps} 
\includegraphics[width=6.0cm,angle=-90]{ESO113-G010_0103861601_multiplot.ps} 
}
\caption{ {\it Left:} The best fit model and the data in the top panel, while the lower two panels are the residuals after the best fit model is employed, for the source ESO~113-G010. } \label{fig:ESO113}
\end{figure*}


\begin{figure*}
  \centering

\vbox{

\hbox{
\includegraphics[width=6.0cm,angle=-90]{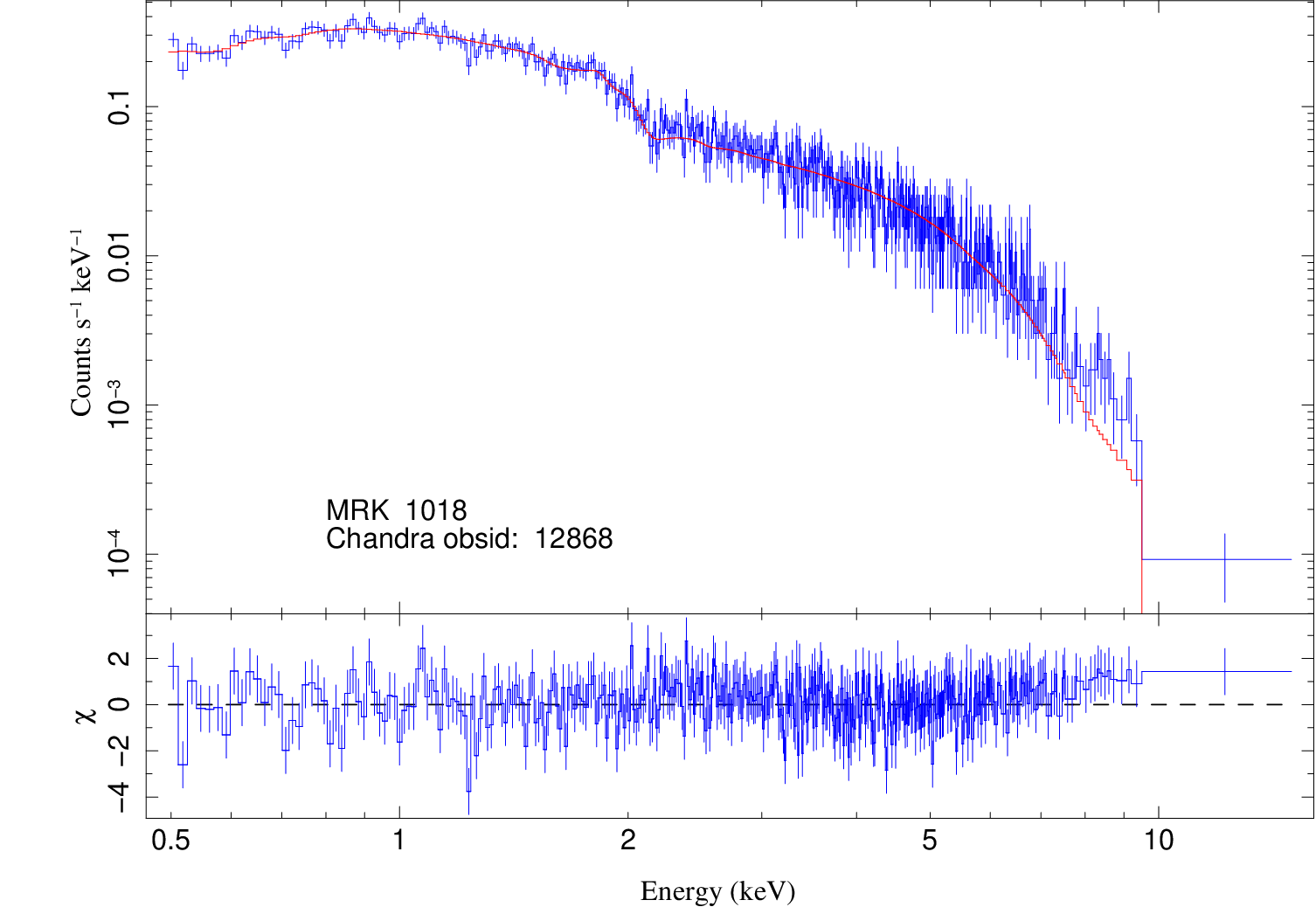} 
\includegraphics[width=6.0cm,angle=-90]{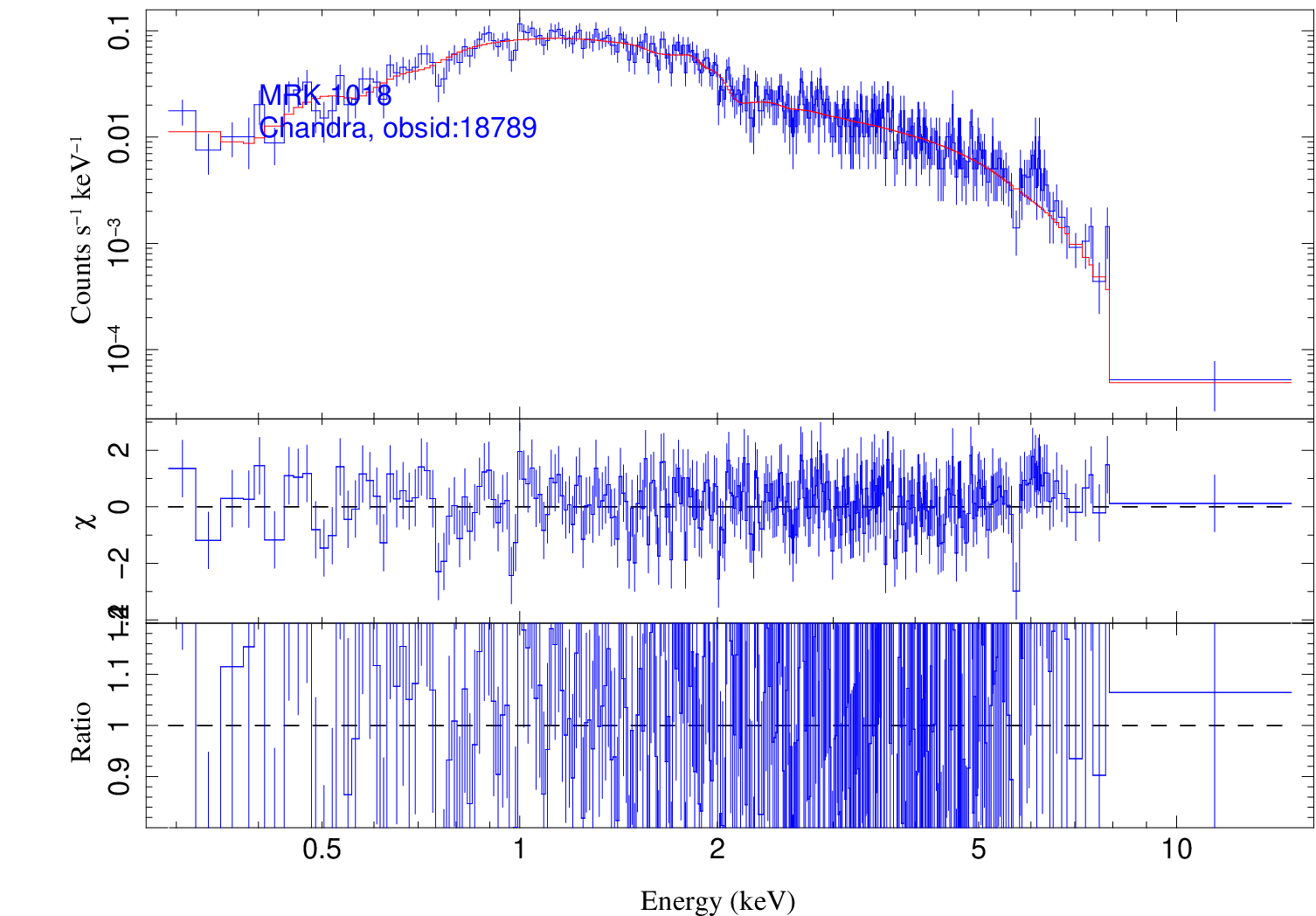} 
}

\hbox
{

\includegraphics[width=6.0cm,angle=-90]{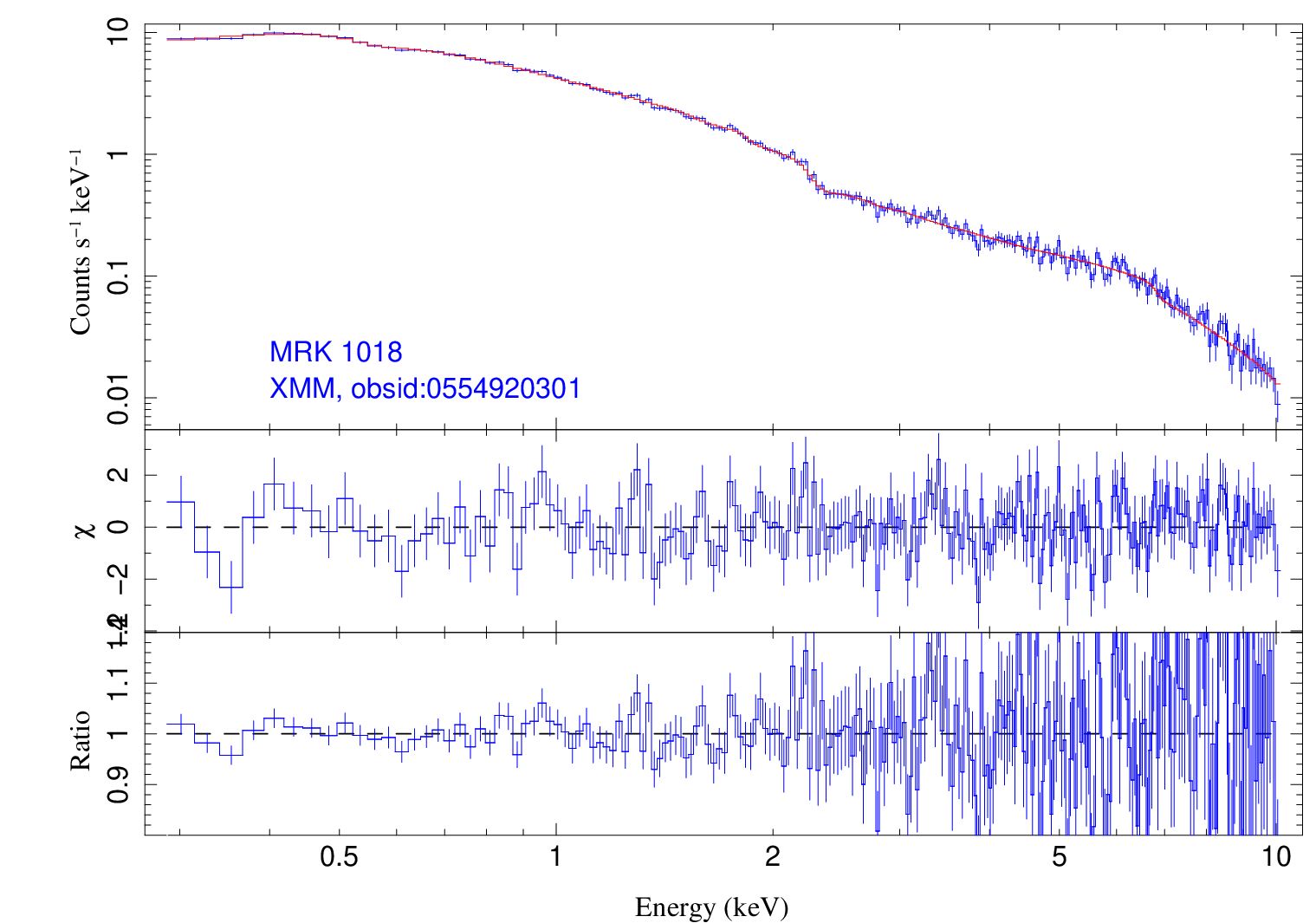}
\includegraphics[width=6.0cm,angle=-90]{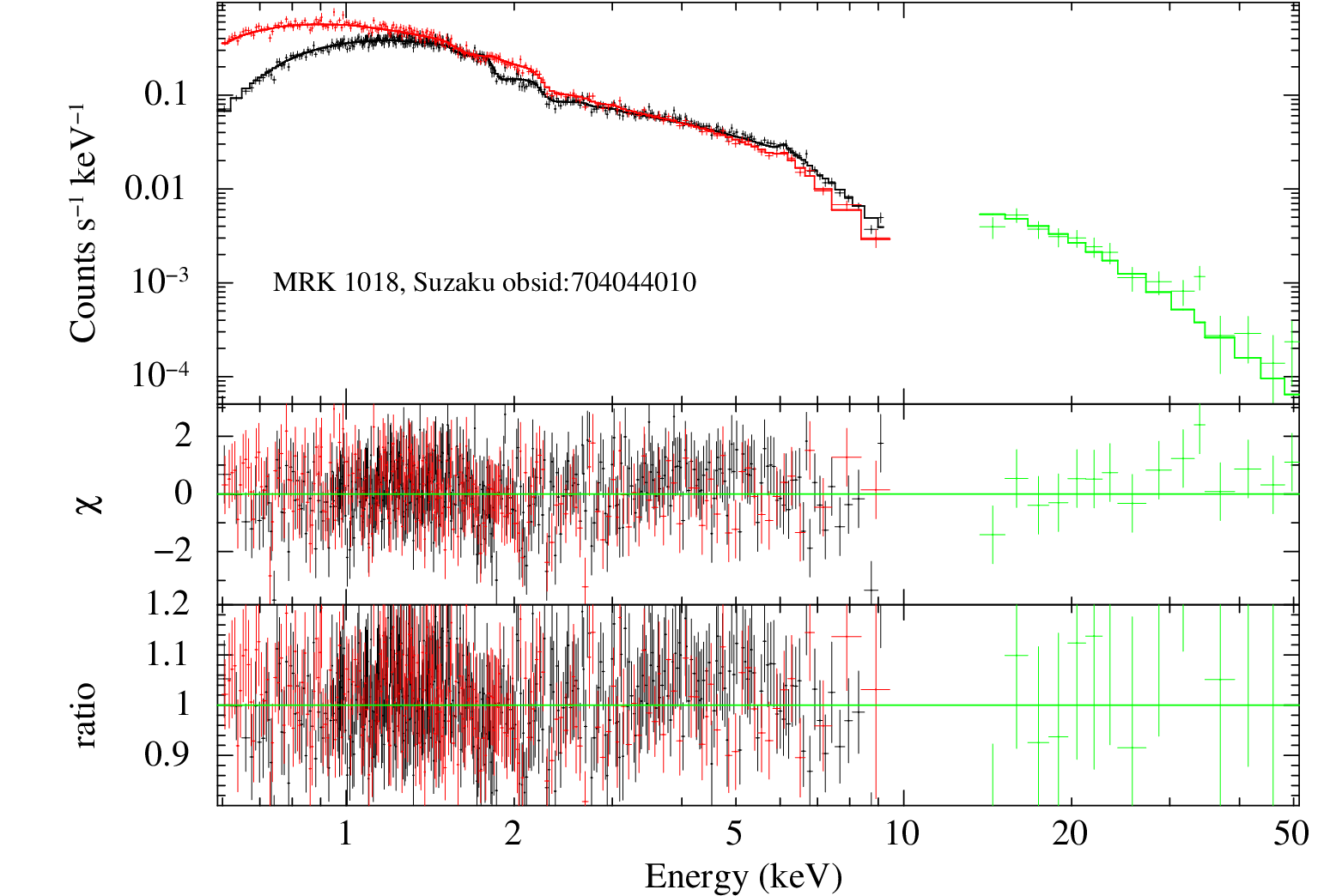}
}

}

\caption{Same as Figure \ref{fig:ESO113}, except for the source which is Mrk~1018.  } \label{fig:MRK1044}
\end{figure*}


\begin{figure*}
  \centering

\vbox{

\hbox{
\includegraphics[width=6.0cm,angle=-90]{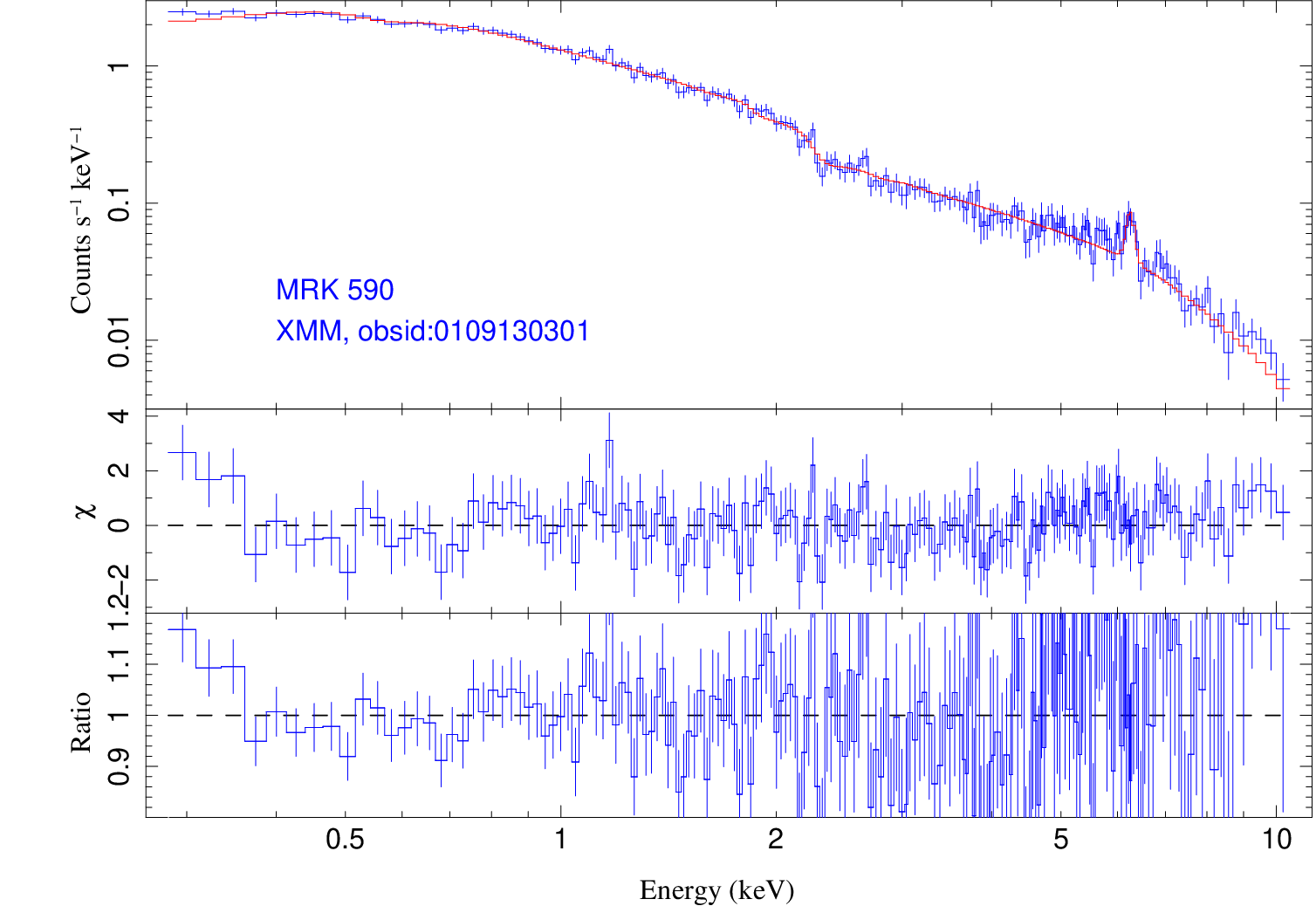} 
\includegraphics[width=6.0cm,angle=-90]{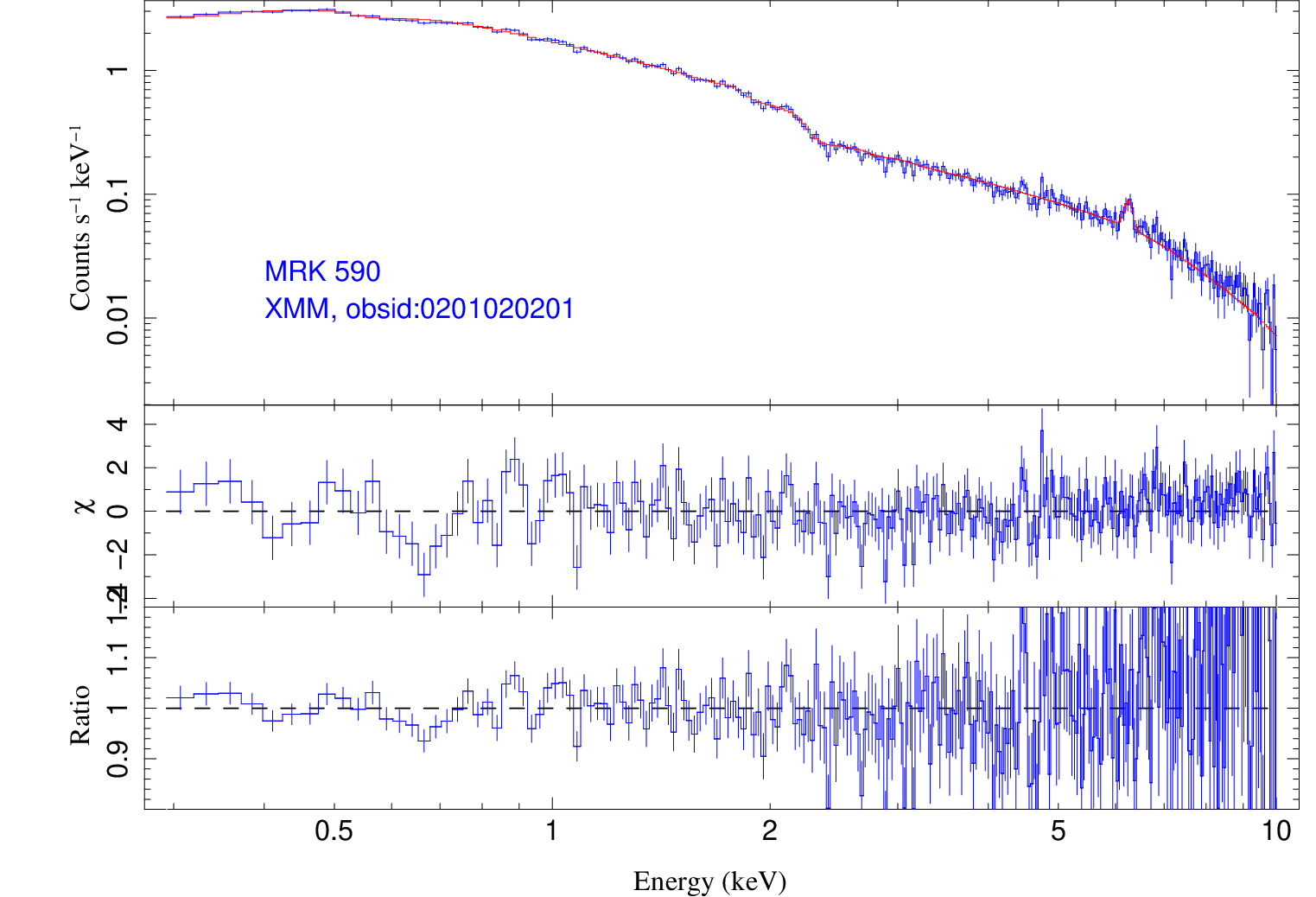} 
}

\hbox{
\includegraphics[width=6.0cm,angle=-90]{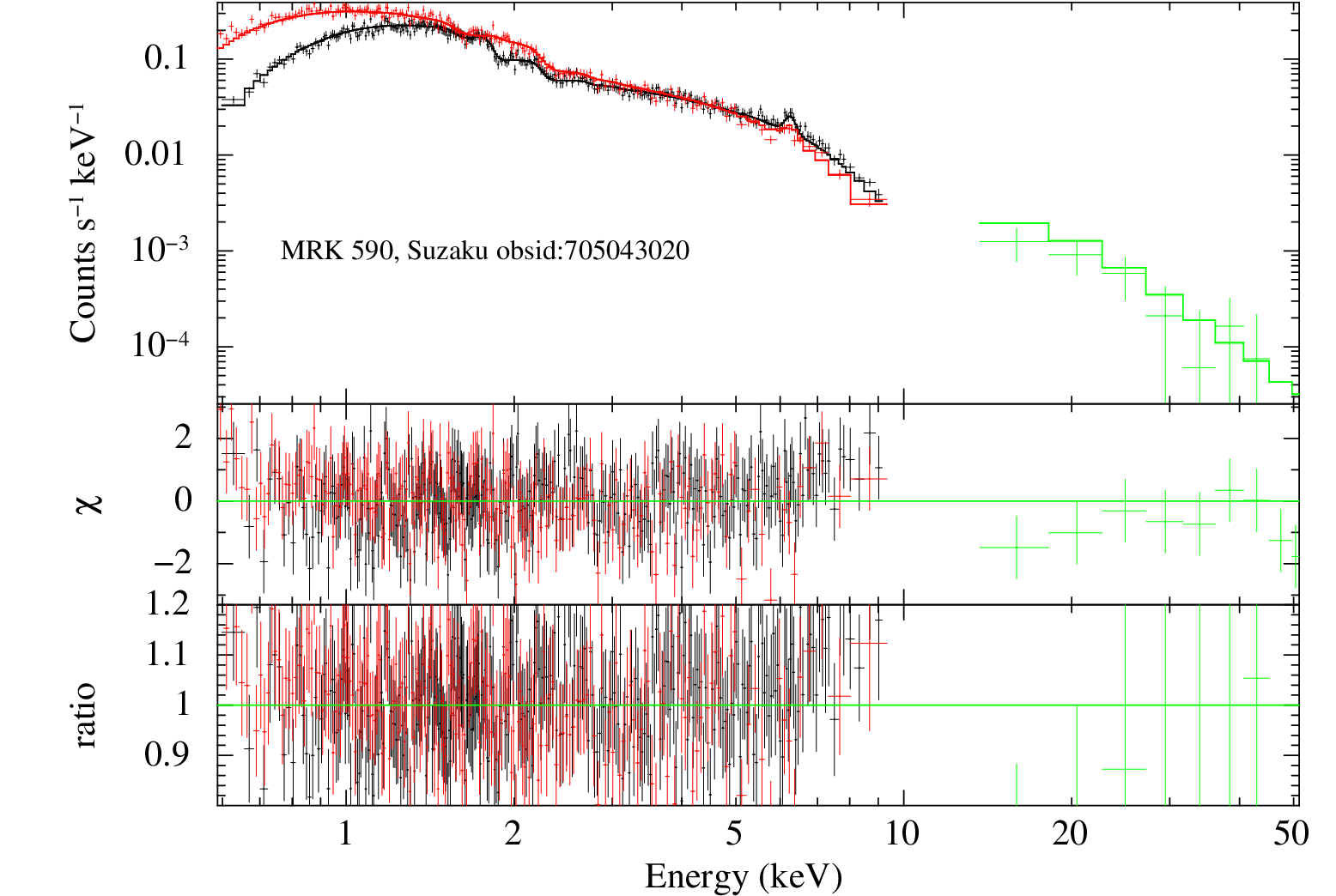} 
\includegraphics[width=6.0cm,angle=-90]{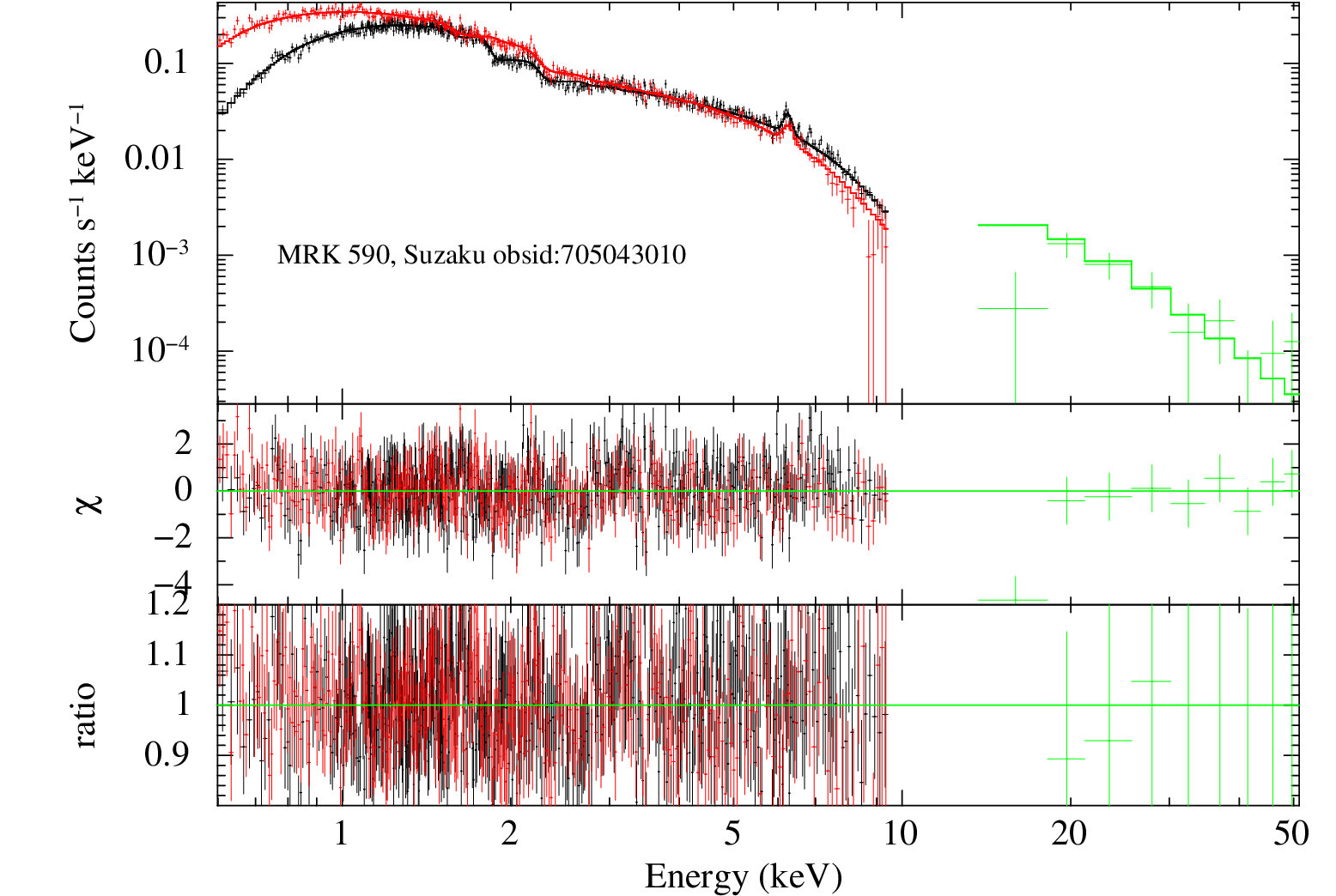} 
}

}
\caption{ Same as Figure \ref{fig:ESO113}, except for the source which is Mrk~590.  } \label{fig:MRK1044}
\end{figure*}


\begin{figure*}
  \centering

\vbox{

\hbox{
\includegraphics[width=6.0cm,angle=-90]{MRK1044_XMM_0112600301_multiplot.ps} 
\includegraphics[width=6.0cm,angle=-90]{MRK1044_XMM_0695290101_multiplot.ps} 
}

\hbox{
\includegraphics[width=6.0cm,angle=-90]{MRK1044_chandra_18685_multiplot.ps} 
}

}

\caption{ Same as Figure \ref{fig:ESO113}, except for the source which is Mrk ~1044. } \label{fig:MRK1044}
\end{figure*}


\begin{figure*}
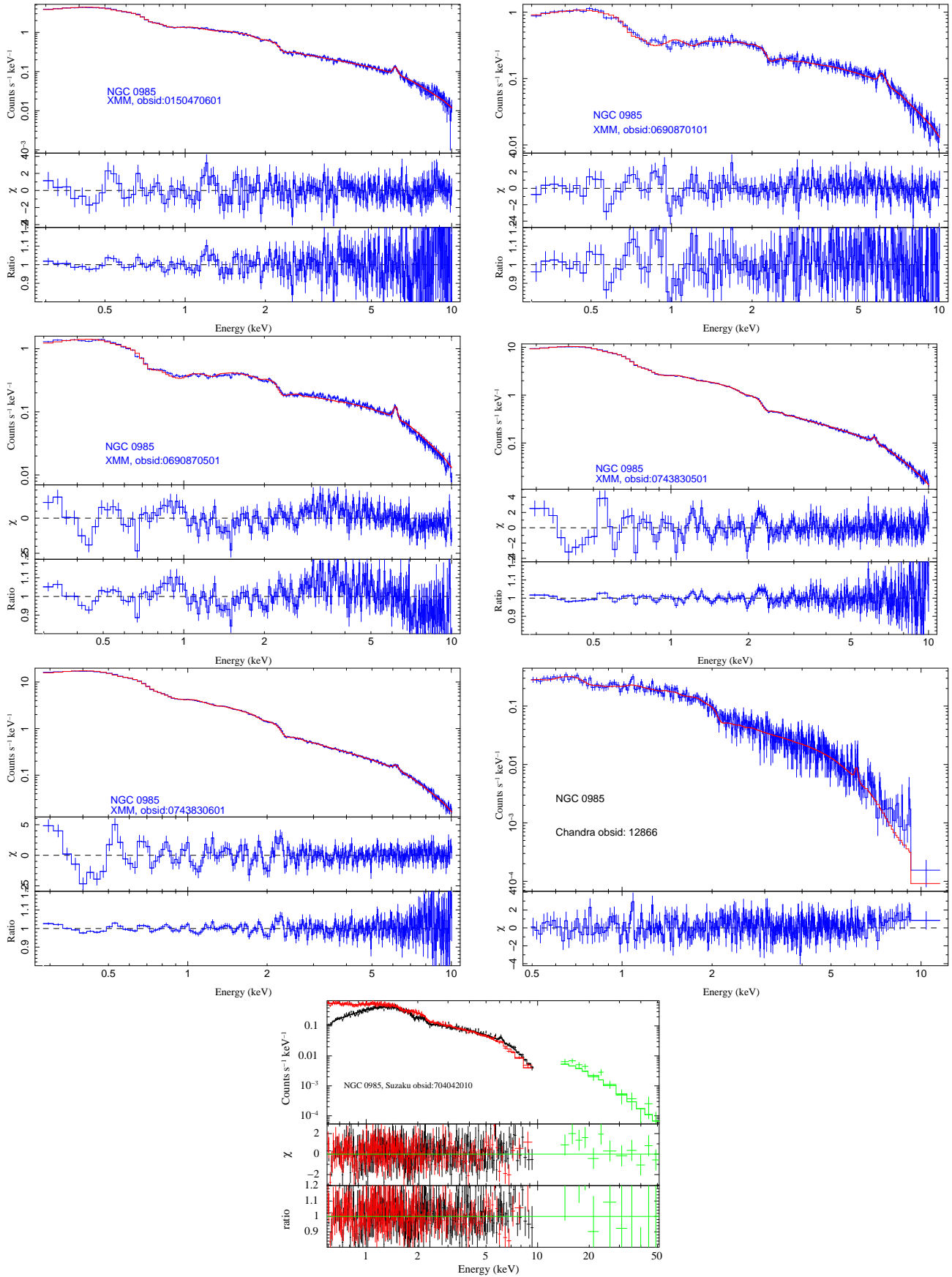

  \centering

\vbox{

\hbox{
\includegraphics[width=6.0cm,angle=-90]{NGC0985_XMM_0150470601_multiplot.ps} 
\includegraphics[width=6.0cm,angle=-90]{NGC0985_XMM_0690870101_multiplot.ps} 
}

\hbox{
\includegraphics[width=6.0cm,angle=-90]{NGC0985_XMM_0690870501_multiplot.ps} 
\includegraphics[width=6.0cm,angle=-90]{NGC0985_XMM_0743830501_multiplot.ps} 
}

\hbox{
\includegraphics[width=6.0cm,angle=-90]{NGC0985_XMM_0743830601_multiplot.ps} 
\includegraphics[width=6.0cm,angle=-90]{NGC0985_chandra_12868_multiplot.ps} 
}

\includegraphics[width=5.0cm,angle=-90]{Best_fit_ztbabs_po_bb_gauss_ngc0985_32ks.ps}

}

\caption{ Same as Figure \ref{fig:ESO113}, except for the source which is NGC~0985. } \label{fig:MRK1044}
\end{figure*}


\begin{figure*}
  \centering 

\includegraphics[width=6.0cm,angle=-90]{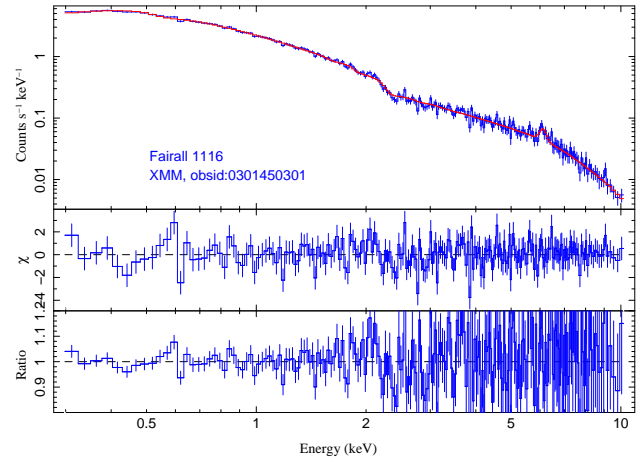} 

\caption{ Same as Figure \ref{fig:ESO113}, except for the source which is Fairall 1116. } \label{fig:MRK1044}
\end{figure*}


\begin{figure*}
  \centering 

	\includegraphics[width=6.0cm,angle=-90]{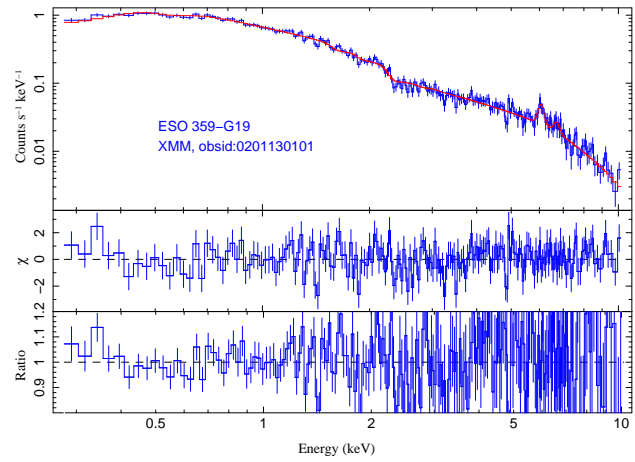} 

\caption{Same as Figure \ref{fig:ESO113}, except for the source which is ESO 359-G19 } \label{fig:MRK1044}
\end{figure*}


\begin{figure*}
  \centering 

\includegraphics[width=6.0cm,angle=-90]{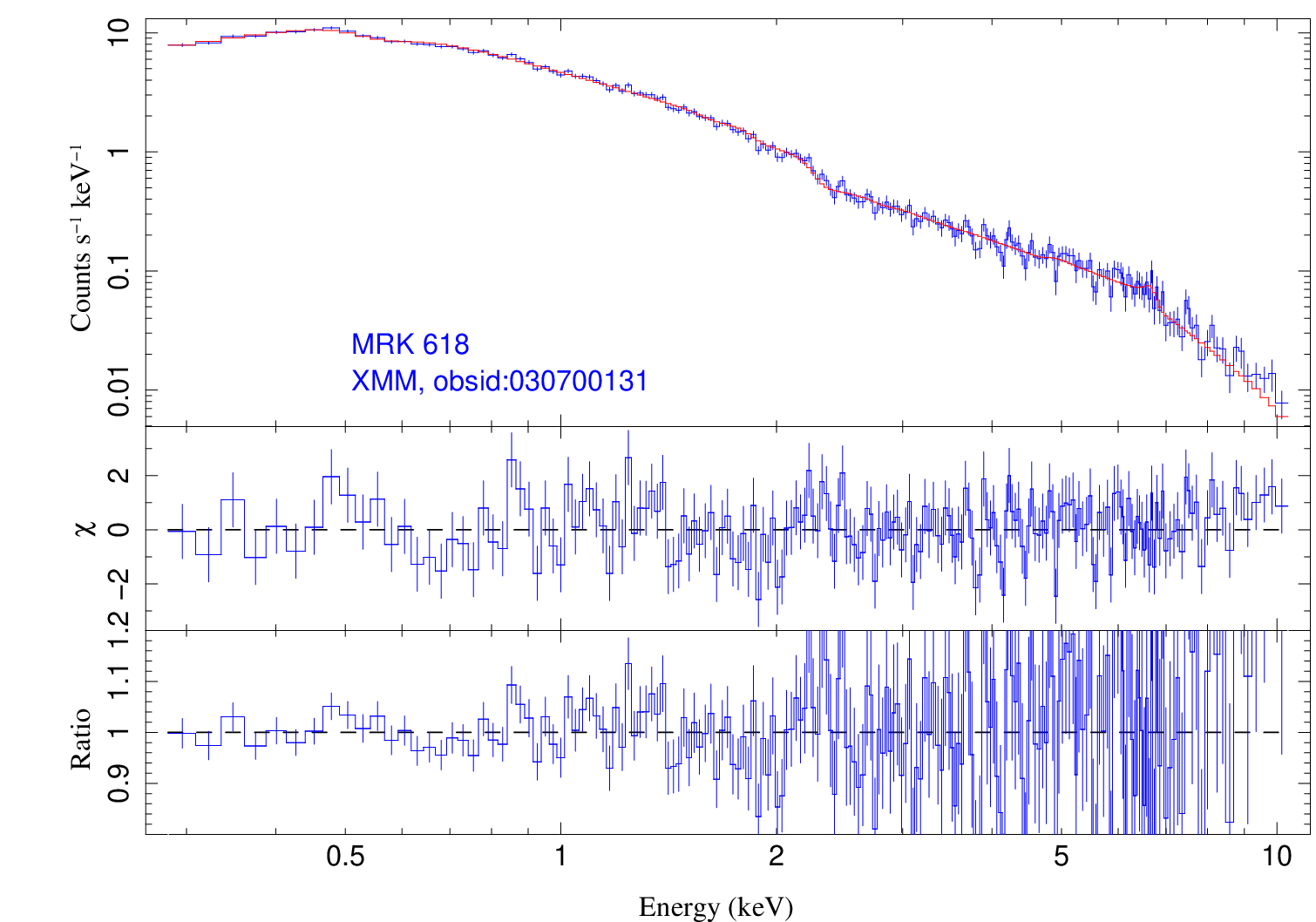} 

\caption{ Same as Figure \ref{fig:ESO113}, except for the source which is Mrk  618. } \label{fig:MRK1044}
\end{figure*}


\begin{figure*}
  \centering 

\hbox{
\includegraphics[width=6.0cm,angle=-90]{MRK1239_XMM_0065790101_multiplot.ps} 
\includegraphics[width=6.0cm,angle=-90]{Best_fit_2cloudy_po_bb_3mytorus_Mrk1239.ps} 
}

\caption{ Same as Figure \ref{fig:ESO113}, except for the source which is  Mrk 1239. } \label{fig:MRK1044}
\end{figure*}


\begin{figure*}
  \centering

\includegraphics[width=6.0cm,angle=-90]{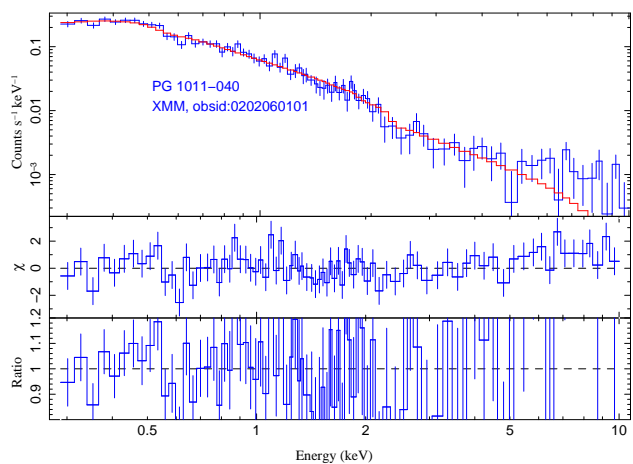} 

\caption{  Same as Figure \ref{fig:ESO113}, except for the source which is  PG 1011-040 } \label{fig:MRK1044}
\end{figure*}


\begin{figure*}
  \centering

\vbox{

\hbox{
\includegraphics[width=5.5cm,angle=-90]{MRK1298_XMM_0202060201_multiplot.ps} 
\includegraphics[width=5.5cm,angle=-90]{MRK1298_XMM_0556230701_multiplot.ps} 
}

\hbox{
\includegraphics[width=5.5cm,angle=-90]{MRK1298_XMM_0556231201_multiplot.ps} 
\includegraphics[width=5.5cm,angle=-90]{MRK1298_XMM_0606150101_multiplot.ps} 
}

\hbox{
\includegraphics[width=5.5cm,angle=-90]{MRK1298_XMM_0728180301_multiplot.ps} 
\includegraphics[width=5.5cm,angle=-90]{MRK1298_XMM_0728180401_multiplot.ps} 
}

\includegraphics[width=5.5cm,angle=-90]{MRK1298_XMM_0728180501_multiplot.ps}

}

\caption{ Same as Figure \ref{fig:ESO113}, except for the source which is Mrk 1298 } \label{fig:MRK1044}
\end{figure*}


\begin{figure*}
  \centering

\includegraphics[width=6.0cm,angle=-90]{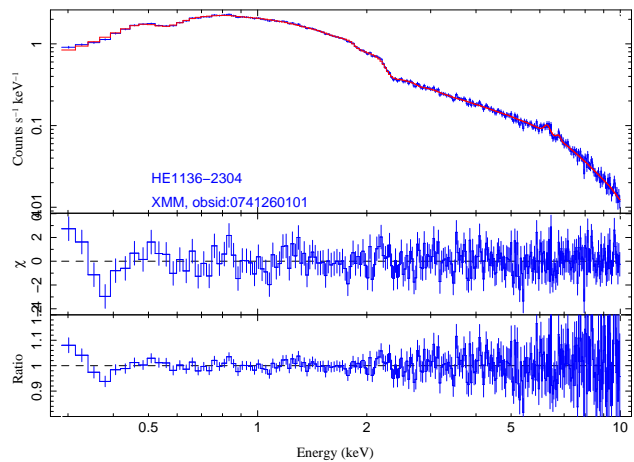} 

\caption{ Same as Figure \ref{fig:ESO113}, except for the source which is  HE 1136-2304. } \label{fig:MRK1044}
\end{figure*}


\begin{figure*}
  \centering 

\hbox{

\includegraphics[width=6.0cm,angle=-90]{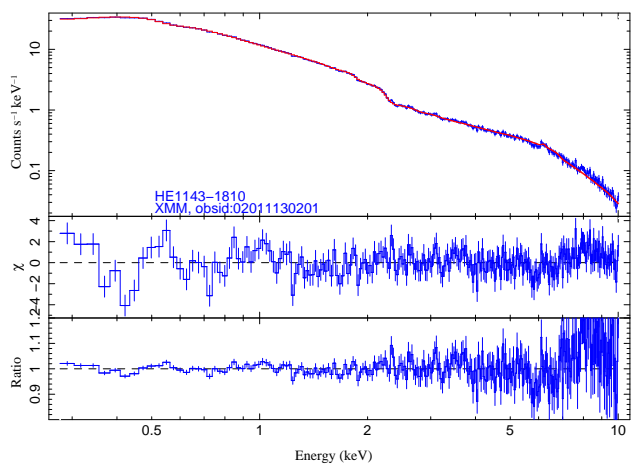} 
\includegraphics[width=6.0cm,angle=-90]{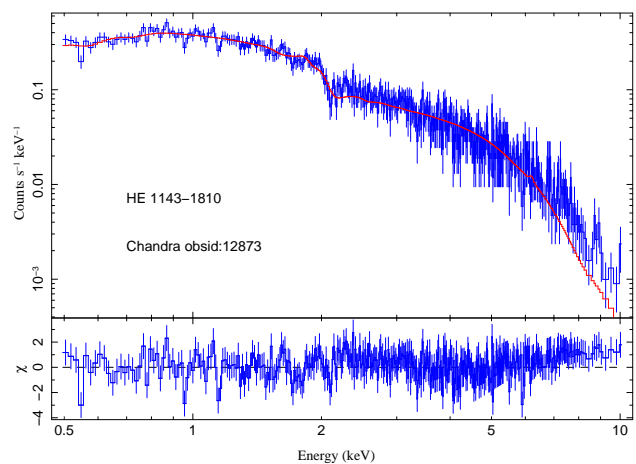} 
}

\caption{Same as Figure \ref{fig:ESO113}, except for the source which is HE 1143-1810 } \label{fig:MRK1044}
\end{figure*}

\clearpage
\begin{figure*}
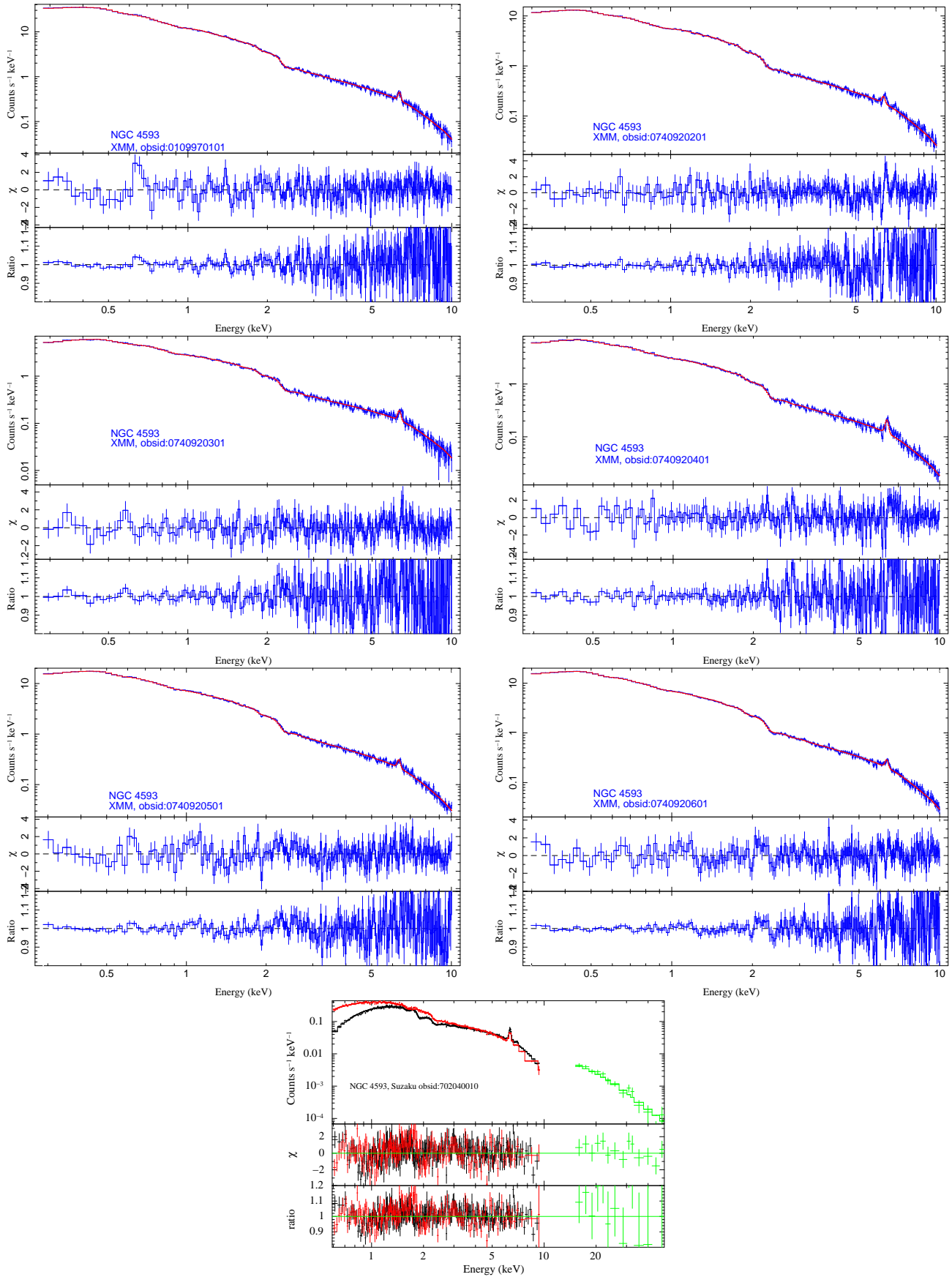

  \centering

\vbox{

\hbox{
\includegraphics[width=6.0cm,angle=-90]{NGC4593_XMM_0109970101_multiplot.ps} 
\includegraphics[width=6.0cm,angle=-90]{NGC4593_XMM_0740920201_multiplot.ps} 
}

\hbox{
\includegraphics[width=6.0cm,angle=-90]{NGC4593_XMM_0740920301_multiplot.ps} 
\includegraphics[width=6.0cm,angle=-90]{NGC4593_XMM_0740920401_multiplot.ps} 

}

\hbox{
\includegraphics[width=6.0cm,angle=-90]{NGC4593_XMM_0740920501_multiplot.ps} 
\includegraphics[width=6.0cm,angle=-90]{NGC4593_XMM_0740920601_multiplot.ps} 

}

\includegraphics[width=5.0cm,angle=-90]{Best_fit_po_bb_2gauss_ngc4593_119ks.ps}

}

\caption{ Same as Figure \ref{fig:ESO113}, except for the source which is NGC~4593 } \label{fig:MRK1044}
\end{figure*}


\begin{figure*}
  \centering

\includegraphics[width=6.0cm,angle=-90]{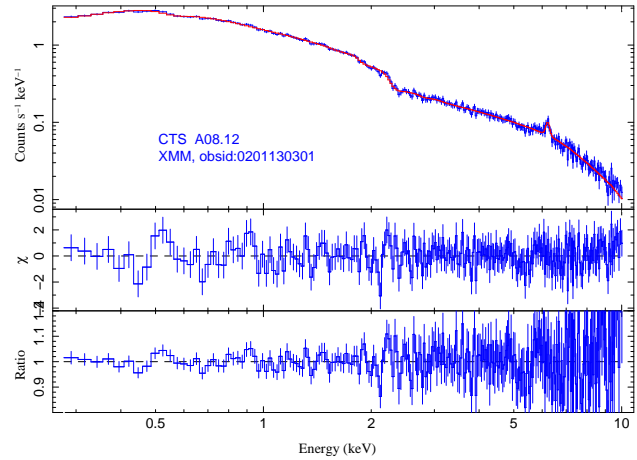} 

\caption{ Same as Figure \ref{fig:ESO113}, except for the source which is CTS A08.12. } \label{fig:MRK1044}
\end{figure*}


\begin{figure*}
  \centering 

\hbox{
\includegraphics[width=6.0cm,angle=-90]{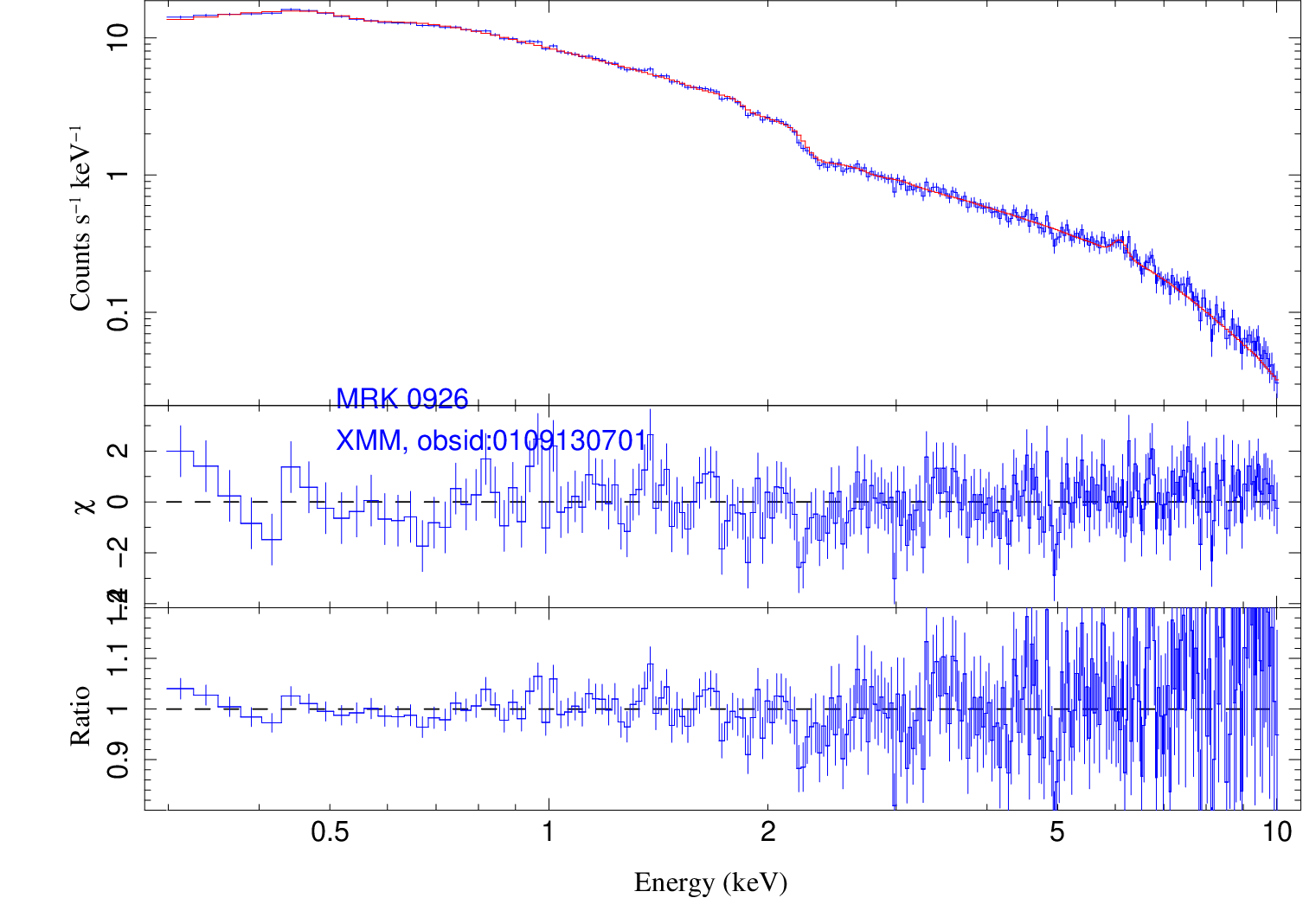} 
\includegraphics[width=6.0cm,angle=-90]{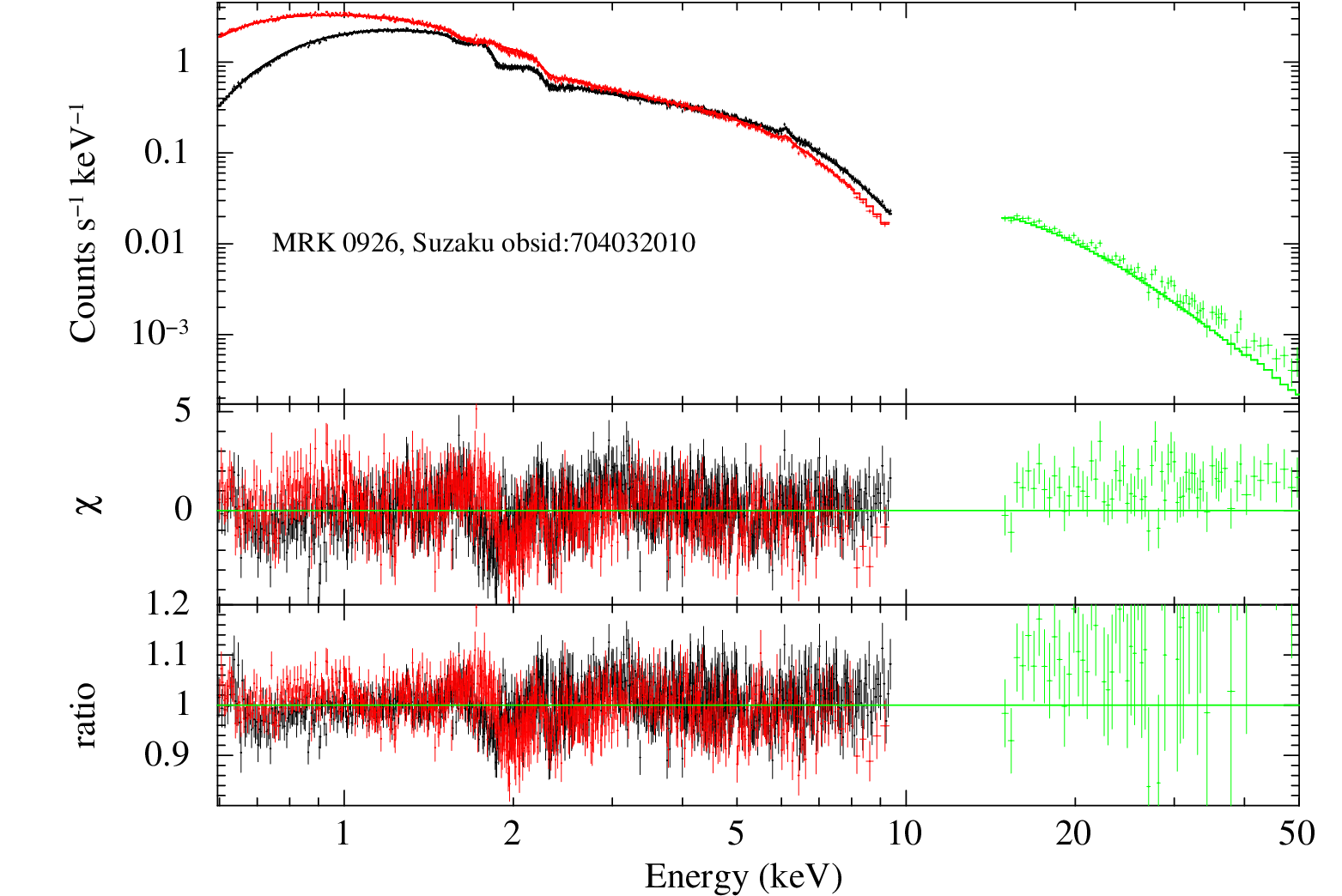} 
}

\caption{ Same as Figure \ref{fig:ESO113}, except for the source which is Mrk 0926. } \label{fig:MRK0926}
\end{figure*}


$Acknowledgements:$  Author SL thanks Gerold Busch for insightful discussions. The authors are grateful to the referee for his/her comments which improved the quality of the paper. This research has made use of the NASA/IPAC Extragalactic Database (NED) which is operated by the Jet Propulsion Laboratory, California Institute of Technology, under contract with the National Aeronautics and Space Administration.

\bibliographystyle{mn2e}
\bibliography{mybib}

\end{document}